\documentclass[final,5p,times,twocolumn]{elsarticle}

\usepackage{amssymb}
\usepackage{lipsum}
\usepackage{graphicx}
\usepackage{comment}
\usepackage[table,xcdraw]{xcolor}
\usepackage{amsmath}
\usepackage{amssymb}
\usepackage{setspace}
\usepackage{hyperref}
\usepackage[nameinlink]{cleveref}
\usepackage{enumitem}
\usepackage{wrapfig}
\usepackage{subcaption}
\usepackage{xurl}
\usepackage{ftnright}
\usepackage[ruled, vlined, linesnumbered]{algorithm2e}
\usepackage{tablefootnote}
\usepackage[T1]{fontenc}
\usepackage{caption}
\usepackage{mathtools}

\let\oldnl\nl
\newcommand{\nlnonumber}{\renewcommand{\nl}{\let\nl\oldnl}}

\SetAlFnt{\footnotesize}
\usepackage{algorithmic}
\crefname{algocf}{alg.}{algs.}
\Crefname{algocf}{Algorithm}{Algorithms}
\usepackage{booktabs}
\usepackage{multirow}

\setlist[description]{leftmargin=\parindent,labelindent=\parindent}
\setlist[itemize]{label=\footnotesize$\bullet$}
\newtheorem{theorem}{Theorem}

\DeclareMathOperator*{\argmin}{\arg\min}
\DeclareMathOperator*{\argmax}{\arg\max}
\newcommand{\pluseq}{\mathrel{+}:=}

\newcommand{\AlgAbrev}{\textsc{FAR}\xspace}
\newcommand{\POORSCALING}{\textsc{PoorScaling}\xspace}
\newcommand{\VARIEDSCALING}{\textsc{MixedScaling}\xspace}
\newcommand{\GOODSCALING}
{\textsc{GoodScaling}\xspace}
\newcommand{\WIDETIMES}
{\textsc{WideTimes}\xspace}
\newcommand{\NARROWTIMES}
{\textsc{NarrowTimes}\xspace}
\newcommand{\MISO}
{\textsc{MISO-opt}\xspace}
\newcommand{\FIXPART}
{\textsc{FixPart}}
\newcommand{\FIXPARTBEST}
{\textsc{FixPartBest}\xspace}

\journal{Journal of Parallel and Distributed Computing, vol. 204 (2025). doi: \href{https://doi.org/10.1016/j.jpdc.2025.105128}{\texttt{10.1016/j.jpdc.2025.105128}}}

\begin{document}

\begin{frontmatter}



\title{Leveraging Multi-Instance GPUs through moldable task scheduling}


\author[ucm]{Jorge Villarrubia\corref{cor1}}
\ead{jorvil01@ucm.es}
\cortext[cor1]{Corresponding author}
\author[ucm]{Luis Costero}
\author[ucm]{Francisco D. Igual}
\author[ucm]{Katzalin Olcoz}
\affiliation[ucm]{organization={Universidad Complutense de Madrid, Departamento de Arquitectura de Computadores y Automática}, 
            city={Madrid},
            postcode={28040},
            country={Spain}}

\begin{abstract}
NVIDIA MIG (Multi-Instance GPU) allows partitioning a physical GPU into
multiple logical instances with fully-isolated resources, which can be dynamically reconfigured. This work highlights the untapped potential of MIG through moldable task scheduling with dynamic reconfigurations. Specifically, we propose a makespan minimization problem for multi-task execution under MIG constraints. Our profiling shows that assuming monotonicity in task work with respect to resources is not viable, as is usual in multicore scheduling. Relying on a state-of-the-art proposal that does not require such an assumption, we present \AlgAbrev, a $3$-phase algorithm to solve the problem. Phase 1 of FAR builds on a classical task moldability method, phase 2 combines Longest Processing Time First and List Scheduling with a novel repartitioning tree heuristic tailored to MIG constraints, and phase 3 employs local search via task moves and swaps. \AlgAbrev schedules tasks in batches offline, concatenating their schedules on the fly in an improved way that favors resource reuse. Excluding reconfiguration costs, the List Scheduling proof shows an approximation factor of 7/4 on the NVIDIA A30 model. We adapt the technique to the particular constraints of an NVIDIA A100/H100 to obtain an approximation factor of 2. Including the reconfiguration cost, our real-world experiments reveal a makespan with respect to the optimum no worse than $1.22\times$ for a well-known suite of benchmarks, and $1.10\times$ for synthetic inputs inspired by real kernels. We obtain good experimental results for each batch of tasks, but also in the concatenation of batches, with large improvements over the state-of-the-art and proposals without GPU reconfiguration. Moreover, we show that the proposed heuristics allow a correct adaptation to tasks of very different characteristics. Beyond the specific algorithm, the paper demonstrates the research potential of the MIG technology and suggests useful metrics, workload characterizations and evaluation techniques for future work in this field.
\end{abstract}




\end{frontmatter}




\section{Introduction}

\subsection{Context and motivation}

Graphics Processing Units (GPUs) are nowadays established as the elemental computing device for accelerating massively parallel computations in fields such as Artificial Intelligence (AI) and High Performance Computing (HPC). Hand in hand  with the emergence of much more resource-demanding tasks, 
manufacturers are offering increasingly powerful GPUs featuring tens of thousands of computing cores and a massive memory bandwidth.
These resources, however, can be difficult to be fully utilised in applications that exhibit 
lower levels of parallelism or bandwidth necessities, resulting in resource underutilization. 
Co-scheduling independent applications is a natural way to regain full occupation of the available computing resources in a single GPU. Still, it poses important challenges to avoid or control resource contention and to adapt the assigned per-application resources to attain fair scheduling policies.

Manufacturers offer different solutions to address the problem of resource isolation 
in multi-tenant scenarios on a single GPU.
For example, the Multi-Processing Service (MPS) from NVIDIA~\cite{mps} allows multiple processes to use the same GPU simultaneously by provisioning the stream multiprocessors (SMs), but sharing memory bandwidth. This is not a good solution for the cloud, as the quality of service a client receives (e.g. processing speed) cannot be guaranteed in advance since it depends on the resource usage of the tasks running simultaneously on the GPU. 
Some of the latest GPU models for data centers, including the NVIDIA Ampere, Hopper and Blackwell families, 
integrate the  Multi-Instance GPU (MIG) technology\footnote{See \url{https://docs.nvidia.com/datacenter/tesla/mig-user-guide/index.html}.}, which allows the GPU to be logically and physically split into disjoint instances that have their own fully isolated hardware resources, including memory bandwidth, cache and cores, among others. 

For Cloud Service Providers (CSPs), MIG solves resource contention in multi-tenant scenarios. 
Typically  deployed in a {\em static} and {\em task-oblivious} fashion 
(that is, the partitioning scheme does not vary frequently and it does not depend
on the specific features of the task(s) to execute), 
the use of MIG ensures that one client cannot impact the 
work of others, providing robust Service Level Agreements (SLA) on a shared GPU without needing to 
deploy tasks across multiple GPUs, and at the same time improving the effective resource utilization. 
However, the reconfigurable nature of MIG-capable GPUs enables {\em dynamic} partitioning and scheduling 
schemes for tasks with different scalability properties. As of today, the joint exploitation of task scheduling and
MIG reconfiguration is a topic hardly covered in the literature.

\subsection{MIG concepts and restrictions}\label{sec:concepts_restriction}

The minimum units in MIG are called {\em slices}, comprising one block of computational resources called GPCs (GPU Processing Clusters), along with a portion of L2 cache and DRAM. Several consecutive slices can be grouped in an {\em instance} on which to execute a task. 
Furthermore, due to physical limitations, it is not possible to form instances with any sequence of consecutive slices. Finally, a set of instances that divide the GPU is called a {\em partition}.

\begin{figure}[ht!]
\centering
\includegraphics[width=\columnwidth] {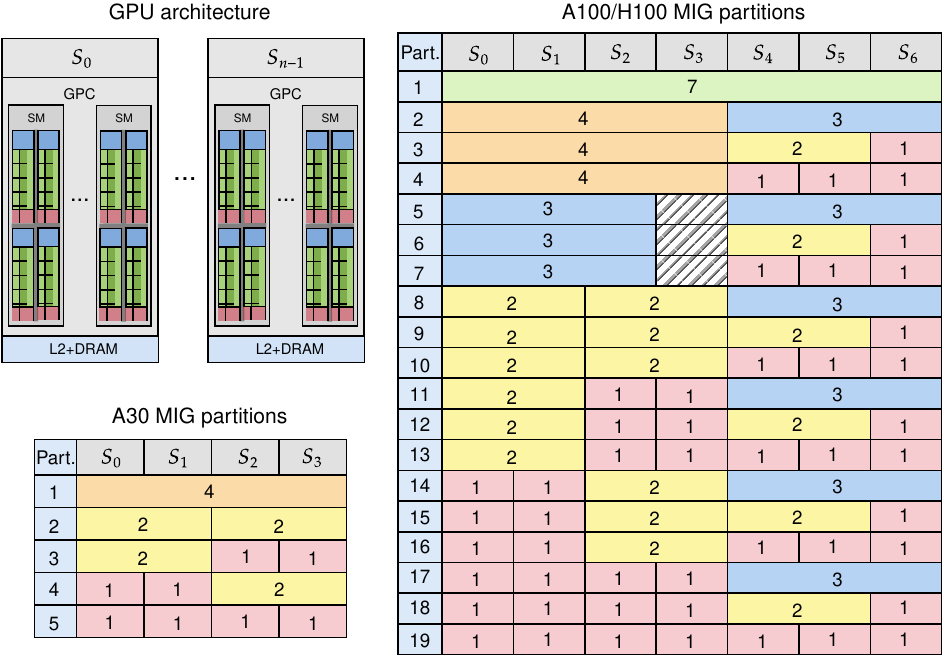}
\caption{Slice per instance distributions for the NVIDIA A30 and A100/H100.}
\label{fig:MIG_A30_A100_partitions}
\end{figure}

\Cref{fig:MIG_A30_A100_partitions} shows all the possible partitions of the NVIDIA A30 (left), consisting of 4 slices with 6 GB of memory assigned to each one. Note that it is not possible to form instances with 3 slices, nor to create an instance with the two middle ones: $\{S_1, S_2\}$. 
The NVIDIA A100 and H100 extend this strategy to 7 slices, with 5 and 10 GB of memory in each slice respectively,
presenting a total of 19 possible partitions, also shown in \Cref{fig:MIG_A30_A100_partitions} (right). 
Note that partitions 5, 6 and 7 do not use $S_3$. In these partitions, the memory allocated to $S_3$ can be assigned to the instance composed by $\{S_0, S_1, S_2\}$, imitating the same instance formed by $\{S_4, S_5, S_6\}$ (as $S_6$ comprises more memory than the rest of slices). 
Except for a slightly lower energy consumption, it never seems more interesting to use $\{S_0, S_1, S_2\}$ disabling $S_3$ than directly using $\{S_0, S_1, S_2, S_3\}$; however, since NVIDIA offers it, we will also consider that instance.

Finally, MIG allows to reconfigure partitions dynamically, so that even if a task assigned to an instance has not completed its execution, other free instances can be repartitioned/merged. This reconfiguration philosophy means that (1) instances can be organized hierarchically (they are joined or split to form other instances), and (2) valid partitions are precisely the possible combinations of disjoint instances. These two properties will guarantee that our proposal works correctly. To make the explanations more concrete and easier to follow, we will base them on the GPUs previously described (NVIDIA A30, A100 and H100), but they can be extrapolated to any future generations as long as they maintain these properties, which is expected. In fact, NVIDIA has just released two models of the Blackwell family (B100 and B200) with exactly the same MIG restrictions as the A100/H100 models: up to 7 slices with the same valid partitions. In addition to this, task migration between instances is not allowed, nor any kind of task preemption.

\subsection{Motivation of MIG usage for task scheduling}


While the amount of available resources in modern GPUs has increased drastically, not all applications can take advantage of these resources, as their degree of parallelism can be insufficient to use all of them, or the communication overhead can limit their speedup. In these situations the use of MIG can improve the resource usage of the GPU without impact in performance. 

\begin{figure}[ht!]
\centering
\includegraphics[width=\columnwidth] {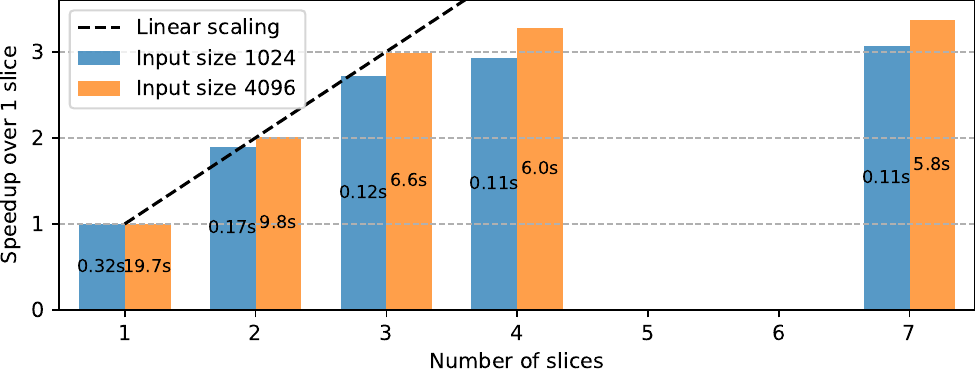}
\caption{Speedup of Rodinia's {\em Gaussian} kernel for instance sizes on A100.}
\label{fig:gaussian_MIG_scaling}

\end{figure}


As an illustrative example, let us consider the {\em Gaussian} kernel within the Rodinia GPU benchmark~\cite{Che2009}, whose performance for an increasing number of slices within an A100 GPU is reported in~\Cref{fig:gaussian_MIG_scaling}. Each color bar represents a different input size, ranging from 0.32s up to 20s for 1 slice (blue and orange bars respectively).
The results reveal that, for this application and both input sizes, it stops scaling beyond 3 slices. In this situation, co-scheduling independent aplications in the
remaining slices would increase the effective GPU occupation, without degrading 
individual performance thanks to the isolation capabilities of MIG.

In addition, in some scenarios, it is common to run the same kernel multiple times, so that all runs can be seen as a longer task for MIG utilization. If the task times are long, the use of MIG makes even more sense as reconfiguration times become negligible, and it is possible that during their executions many new pending jobs accumulate.
%
%
This is specially interesting in datacenters or clusters where typically large tasks are executed. In these scenarios, where many new jobs accumulate in request queues while other long tasks are running, it would be possible to schedule them offline to optimize their joint execution through MIG. 
As of today, MIG is supported in job management systems like SLURM or HTCondor, and in 
workload orchestrators like Kubernetes; their scheduling policies, however, do not include 
coupled strategies for dynamic MIG reconfiguration and job scheduling.

\subsection{Contributions}

In this paper, we address the scheduling of independent tasks (applications) on a MIG-capable GPU, determining the instance and the time at which their execution starts. 
We focus our proposal towards a scenario in which we can form successive batches of pending tasks while executing other long tasks (e.g. taking a batch of tasks every periodic time interval), as is common in batch job management systems and orchestrators. 
For each batch, we consider an {\em offline} scheduling problem (its tasks are available from the start of batch execution), with the objective of minimizing the makespan (last completion time of a task in the batch). This scheduling is {\em moldable}, since the amount of resources that each task receives is not given, but decided by the scheduler upon task start, and it is restricted to always use feasible partitions according to the rules imposed by MIG. Between batches, we perform an {\em online} scheduling (batches are known on the fly), by concatenating their schedules in an {\em intelligent} way. Our proposal is suitable for any GPU where MIG still satisfies the properties mentioned in \Cref{sec:concepts_restriction}: (1) instances can be hierarchically organized, and (2) valid partitions are precisely combinations of disjoint instances. Because of the dynamic and instance-independent reconfiguration philosophy, MIG is expected to preserve these conditions.
%

We assume that the execution time of a task on each instance size is known\footnote{The time only depends on the number of slices in the instance, since the slices are identical. The exception is $S_6$ in the A100, which offers more memory to instances of size 3, but we explained that they also get it in other cases.}, as tasks can be previously profiled to predict their execution time (for example, on an auxiliary node) \cite{Amarís2016A,Johnston2018}, or we operate in a private environment with tasks that have known behaviors and are easily predictable. In fact, we could use the MISO predictor presented in \cite{Li2022}, which precisely predicts execution times for MIG with different instance sizes using profiling-based machine learning.


The main contributions of the paper, together with the section of the manuscript in
which they are covered, are:

\begin{itemize}[itemsep=2pt]
    \item We formulate a scheduling problem for MIG that, to the best of our knowledge, is new. We also review similarities and differences with other problems addressed in the literature,
    showing experimentally that it is not acceptable to assume monotony in the work of the tasks, as most of the moldable scheduling proposals do, although it is reasonable to assume that their times are monotonic.
     %
    This contribution is covered in \Cref{sec:scheduling_problem}.
    
    \item We propose \AlgAbrev ({\small \em Family of Allocations and Repartitioning}), a three-phase algorithm for solving the offline problem for each batch of tasks. The first phase is a direct application of the molding method presented in \cite{Turek92}. The second phase combines Longest Processing Time First and List Scheduling \cite{Graham1969} with a new repartitioning tree heuristic derived from MIG constraints. Finally, the third phase refines the result through task moves and swaps. The details of the algorithm are given in \Cref{sec:algorithm}. A C++ implementation for MIG-capable GPUs is offered to the community in \cite{far_implementation}.
    
    \item We propose a strategy to concatenate \AlgAbrev schedules for multiple task batches to favor resource utilization. This strategy is developed in \Cref{sec:multi_batch}.
    
    \item We applied the List Scheduling proof to derive an approximation factor of 7/4 for the A30 model, without considering reconfiguration costs. We adapted the proof to the casuistry of the A100/H100 models proving an approximation factor of 2 for them. This proof is developed in \Cref{sec:error_bounds}. 
    
    \item We empirically test our proposal, verifying that it provides good solutions, with differences to the optimum of less than 10\%, and does not require large task batches to achieve it. In particular, we show large improvements over other approaches: between 55\% and 230\% versus a scheduler without reconfiguration and between 19\% and 114\% versus the state-of-the-art in MIG scheduling \cite{Li2022}. We use real execution times from Rodinia, as well as synthetic times that mimic real task behaviors and allow adjusting different workloads parametrically.
    The experimental evaluation corresponds to \Cref{sec:experimental}.
\end{itemize}


\section{Scheduling problem\label{sec:scheduling_problem}}

\subsection{Reconfiguration times}\label{sec:reconfig_times}

Reconfiguration times in MIG can greatly influence on the final execution time depending on their
number and cost. 
The reconfiguration of a MIG partition is performed by the isolated destruction and creation of instances; for example, to transform partition 2-2 (partition number 2 for the NVIDIA A30 in \Cref{fig:MIG_A30_A100_partitions}) into partition 1-1-2 (partition number 4 for the NVIDIA A30 in \Cref{fig:MIG_A30_A100_partitions}) it is enough to destroy $\{S_0,S_1\}$, and create $\{S_0\}$ and $\{S_1\}$, without modifying $\{S_2, S_3\}$. Note that this way of reconfiguring by creating/destroying instances independently supports the properties that our proposal will use: (1) hierarchical organization of instances, and (2) any feasible combination of instances is part of a valid partition.

It would be expected that creating/destroying multiple independent instances could be parallelised, but our experience is that the NVIDIA drivers currently sequentialize this process. On the three currently available MIG-capable GPUs (A30, A100 and H100) we have obtained virtually the same reconfiguration time between any pair of partitions by running all builds/destructions sequentially versus using different threads to perform in parallel those that are independent.

\begin{table}[ht!]
\centering

\setlength{\tabcolsep}{3pt}

\caption{Average time (in seconds) for the creation/destruction of an instance of each possible size on each MIG-capable GPU.}
\label{tab:reconfig_times}

\small
\setlength{\tabcolsep}{2pt}
\begin{tabular}{ccccccccc}\toprule
\multirow{2}{*}{\shortstack{Instance\\Size}} & \multicolumn{2}{c}{A30} && \multicolumn{2}{c}{A100} && \multicolumn{2}{c}{H100} \\\cmidrule{2-3}\cmidrule{5-6}\cmidrule{8-9}
                                             & Create & Destroy && Create & Destroy && Create & Destroy \\\midrule
1 & 0.11 & 0.10 && 0.16 & 0.20 && 0.16 & 0.21\\
2 & 0.12 & 0.10 && 0.17 & 0.20 && 0.21 & 0.23\\
3 & --   & --   && 0.20 & 0.21 && 0.33 & 0.25\\
4 & 0.13 & 0.10 && 0.21 & 0.21 && 0.38 & 0.26\\
7 & --   & --   && 0.24 & 0.22 && 0.42 & 0.26\\ \bottomrule
\end{tabular}%
\end{table}

Therefore, the reconfiguration overhead is determined by the sum of the instance creation/destruction times which only depend on their size, as we have also experienced (e.g., it took practically the same time to build/destroy $\{S_0, S_1\}$ as $\{S_2, S_3\}$ because both have size 2). \Cref{tab:reconfig_times} reports the times of each operation on the MIG capable-GPUs for the different instance sizes. Deviations are not reported because in all cases they are of the order of milliseconds, i.e., none of the decimals shown change. We can see that the times are small in all cases, and therefore the reconfigurations should have little impact relative to the duration of the tasks unless they are very short. Reconfigurations were carried out using the NVML library\footnote{\url{https://docs.nvidia.com/deploy/nvml-api/group__nvmlMultiInstanceGPU.html}}, which does not introduce the overhead observed when using the {\tt nvidia-smi} tool. The times in \Cref{tab:reconfig_times} are practically stable between different executions of the same operation, and we will use them for the problem specification and in our scheduling algorithm.

\subsection{Specification and notation}\label{sec:formalisation}

Given a set of independent tasks to be executed on a MIG-capable GPU denoted by $G$ \big(currently, $G \in \{\text{A30}, \text{A100}, \text{H100} \}$\big), we assume that tasks are grouped in batches periodically
\footnote{Long tasks favor a correct batching according to these strategies, but nothing prevents to use some online heuristic for cases with few accumulated tasks.}. For each batch, we pose the following problem:

\begin{description}[itemsep=-2pt]
    \item[\textbf{Input:}] A batch of tasks $\big\{T_1, \dots , T_n\}$ to be executed with MIG on $G$, and their execution times for its instance sizes:
    \begin{gather*}
        t_i: C_G \rightarrow \mathbb{R}^+\;\; \forall\, i \in \{1,\dots,n\},\\
        \text{ where } C_G= \big\{\,|\mathcal{I}| : \mathcal{I} \in \text{MIG-Instances}_G\}.
    \end{gather*}
    In our case, $C_{\text{A30}}\!=\!\{1,2,4\}$ and $C_{\text{A100}}\!=\!C_{\text{H100}}\!= \!\{1,2,3,4,7\}$.\footnote{The number of slices and instance sizes $C_G$ are constant for a particular GPU. However, we include them in the complexity analyses, emphasizing that they do not matter in asymptotic terms and currently have little impact on cost.}
    
    The destruction/creation times of each size are also known:
    \begin{gather*}
        t_{destroy},\, t_{create}: C_G \rightarrow \mathbb{R}^+.
    \end{gather*}
    \item[\textbf{Goal:}] Assign to each task $T_i$ an instance $\mathcal{I}_i \in \text{MIG-Instances}_G$ and a begin time $b_i \geq 0$, trying to minimize the makespan (last finish time $f_i$) without preemption:
    \begin{equation*}
        \min_{\{(\mathcal{I}_i,\,b_i)\}_{i=1}^n}\!\!\Big(\max_{1 \leq i \leq n} f_i \Big),\; \text{ where } f_i = b_i+t_i(|\mathcal{I}_i|).
    \end{equation*}

    \item[\textbf{Constraints:}]
    \
    \begin{enumerate}[leftmargin=*, itemsep=0pt]
        \item Each resource can only be in use by one task, i.e., two tasks with slices in common do not run at the same time:
        \begin{equation*}
            \big[b_i, f_i\big)\,\cap\,\big[b_j, f_j\big) = \varnothing,\quad \text{ if } \mathcal{I}_i\cap\mathcal{I}_j \neq \varnothing,\,\, 1 \leq i \neq j \leq n.
        \end{equation*}
        \item Let $P_t\!=\!\left\{\mathcal{I}_i : t \in \big[b_i, f_i\big)\right\}$ be the set of instances used at time $t$, and $\text{MIG-Part}_G$ the set of valid partitions for $G$ (those shown in \Cref{fig:MIG_A30_A100_partitions}). At any time, the running instances are part of a valid MIG partition:
        \begin{gather*}
            \forall\,t \in \mathbb{R}^+\,\exists P \in \text{MIG-Part}_G,\quad \text { with } P_t \subseteq P.
        \end{gather*}
        \item
        There is enough time for reconfiguration, i.e., between two instants $t < t'$ there is time to sequentially destroy the instances that have disappeared ($\mathcal{I} \in P_t,\,\mathcal{I} \notin P_{t'}$) and create the instances that have appeared ($\mathcal{I}' \notin P_t,\, \mathcal{I}' \in P_{t'}$):
        \begin{equation*}
            \forall\,t < t' \in \mathbb{R}^+, \quad t + \sum\limits_{\mathclap{\mathcal{I} \in P_t,\,\mathcal{I} \notin P_{t'}}} t_{destroy}(|\mathcal{I}|) + \sum\limits_{\mathclap{\mathcal{I}' \notin P_t,\, \mathcal{I}' \in P_{t'}}} t_{create}(|\mathcal{I}'|) \leq t'.
        \end{equation*}
    \end{enumerate}
\end{description}

The schedule of a batch is planned after that of the previous batch, but it is possible to overlap part of their executions while complying the constraints. \Cref{sec:multi_batch} will show how to optimize this concatenation of batch schedules to maximize such overlap.

\subsection{Related work}

Parallel task scheduling has been extensively studied in the literature for different levels of flexibility \cite{Drozdowski1996,Feitelson1997}. In the {\em rigid} form, the amount of resources allocated to each task is predefined and given as input \cite{Bampis2002,Jansen2012}. In the {\em moldable} form, the scheduler statically determines the amount of resources for each task, which remains fixed during execution \cite{Mounie2007,Wu2023}. That is what happens for MIG. Finally, in the {\em malleable} form, the scheduler dynamically adjusts the number of resource allocated, varying them during  task execution \cite{Blazewicz2006,suscom_costero,Marchal2018}. 
%
Additionally, the problem can be {\em offline} if a set of tasks is available from the beginning \cite{Abdelzaher1999,Blazewicz2001,Mounie2007}, or {\em online} if tasks are received and scheduled individually on the fly \cite{Perotin2024,Shmoys1995,Ye2018}. Although the online form is more flexible and suitable for real-time scenarios, it is poorly informed and its results are much worse than the offline equivalent \cite{Guerreiro2023}. In our scenario, where tasks are long enough to form batches, offline scheduling is preferable. 

Several papers have explored scheduling problems involving setup times and costs, with theoretical contributions relevant to dynamic and even stochastic environments \cite{Allahverdi2022,Allahverdi2015}, but most are practical applications in concrete scenarios \cite{Abreu2020,Toksari2022}. Although these approaches offer some flexibility in handling diverse environments, their applicability to the rigid partitioning constraints of MIG is not yet possible. Some authors consider also a bi-criteria objective problem, where another metric is tried to be optimized together with the makespan~\cite{Dutot2004,Sudarsan2016}. Finally, many papers deal with the problem under task precedence constraints given as a graph \cite{Benoit2023,Marchal2018,Perotin2021}.

As discussed above, MIG presents an opportunity via \emph{moldable scheduling} as the number of slices for a task can be chosen at the beginning but cannot be changed until it is finished. In this case, it is usual to approach the problem in two phases: first, by allocating the number of resources that will be available for each task, and second, by applying an algorithm for the resulting problem~\cite{Perotin2024}. Our proposal will work like this, and will treat the rigid problem as the Strip Packing problem~\cite{Baker80}, since they are equivalent when the allocated resources are imposed to be consecutive (this happens with MIG). Due to its strong NP-hardness~\cite{Du89}, this problem has been treated with bio-inspired algorithms~\cite{Bortfeldt2006,Gomez-Villouta2010}, relaxed linear programming models~\cite{Floudas2005,Tarplee2015}, machine-learning approaches~\cite{Fang2023} and, mostly, using heuristics~\cite{Harren2009,Steinberg97}. Among them, we adapt the list scheduling heuristic with LPT (Long Processing Time first)~\cite{Graham1969} for MIG, due to its simplicity and suitability for our problem.

Nevertheless, to the best of our knowledge, there is no version of the problem with constraints such as those imposed by MIG: the set of valid partitions is not the total set of partitions. Thus, using previous approaches to our problem could create partitions such as 2-4-1 in the A100/H100, when the valid one would be 4-2-1 (see~\Cref{fig:MIG_A30_A100_partitions}). Although in such a case one may think of simply reordering the instances to make the partitioning valid, this breaks the inter-task placement intended by the scheduler and tends to produce very poor results. Moreover, none of the existing proposals allows to easily introduce MIG constraints to astutely schedule tasks. 

In practice, most proposals focus on multicores and assume a task monotony hypothesis that, as we will see shortly, is not acceptable on MIG-capable GPUs. The problem is easier to solve under the monotony assumption since it becomes NP-hard in the weak sense, and polynomial algorithms emerge to arbitrarily approximate optimal scheduling \cite{Jansen2018}. In fact, for task execution times it is usual to assume a more restrictive scaling model, such as Roofline, the communication model or Amdahl's model, which in particular satisfy the monotony hypothesis \cite{Benoit2023}. In contrast, approaches for non-monotonic tasks are forced to relax the problem in another way to approximate arbitrarily well the optimal solution: by assuming a constant number of processors \cite{Amoura2002,Jansen2002}, or a polynomial number of processors with respect to the number of tasks \cite{Jansen2010}. Although these relaxations are adequate in our case (the number of slices is constant and, in fact, small), the cost hides excessively large multiplicative constants and degrees that limit their practical use. In this sense, it is preferable to use much more efficient heuristic proposals, even if they do not guarantee such a good approximation \cite{Hunold2015,Turek92}. Among the few proposals in this line we highlight the {\em family of allotments} presented by Turek et al. \cite{Turek92}, which is oriented to independent tasks and will be used for phase 1 of our scheduler.

Finally, there are not many proposals of scheduling with MIG, probably due to its novelty. The aim of \cite{Tan2021} is to minimize the number of GPUs satisfying the QoS requirements of the tasks. Instead, MISO optimizer presented in \cite{Li2022} iteratively selects the partition that maximizes the sum of the speedups of tasks in FIFO order. 
Although we previously highlighted the time predictor of that paper, we believe that their scheduling optimizer has room for improvement. We will experimentally compare our scheduler against it in \Cref{sec:experimental}. Very recently, some proposals aiming at improving the use of MIG in multi-tenant scenarios have been published. Among them, ElasticBatch \cite{Qi2024} employs a machine learning based pipeline to reduce the number of GPU instances and increase their utilization. Moreover, Jormungandr \cite{Wei2024} improves the use of MIG for the particular case of deep neural networks (DNNs) with a containerized implementation using Kubernetes. Finally, MIGER~\cite{Zhang2024} combines the use of MIG and MPS to improve the performance of offline task submission with QoS requirements.

\subsection{Non-monotonic tasks}\label{sec:no-monotony}
Most proposals for moldable scheduling include a task monotony hypothesis consisting of the following two points:
\begin{enumerate}
    \item A task takes the same or less time to execute when it has more resources:
    \begin{equation*}
        t_i(s) \geq t_i(s'),\quad \forall\, 1 \leq i \leq n,\, s < s'.
    \end{equation*}
    \item The work of a task (product of the amount of resources by its execution time) does not decrease with more resources. This means that no task scales super-linearly:
    \begin{gather*}
        s \cdot t_i(s) \leq s' \cdot t_i(s')\equiv \text{speedup}_i(s)\,/\,s \geq \text{speedup}_i(s')\,/\,s',\\ \forall\, 1 \leq i \leq n,\, s < s', \\
        \text{where } \text{speedup}_i(m) = t_i(1)\,/\,t_i(m).
    \end{gather*}  
\end{enumerate}


The second point, which is of capital importance to take advantage of the hypothesis in algorithmic terms, assumes linear scaling as ideal and conceives sub-linear scaling by overhead in communications. However, sometimes it is possible to observe super-linear scaling due to memory hierarchy effects, the increasing available bandwidth or scheduling anomalies such as those described in \cite{Graham1969}.

\begin{figure}[ht!]
\centering
\includegraphics[width=0.47\textwidth]{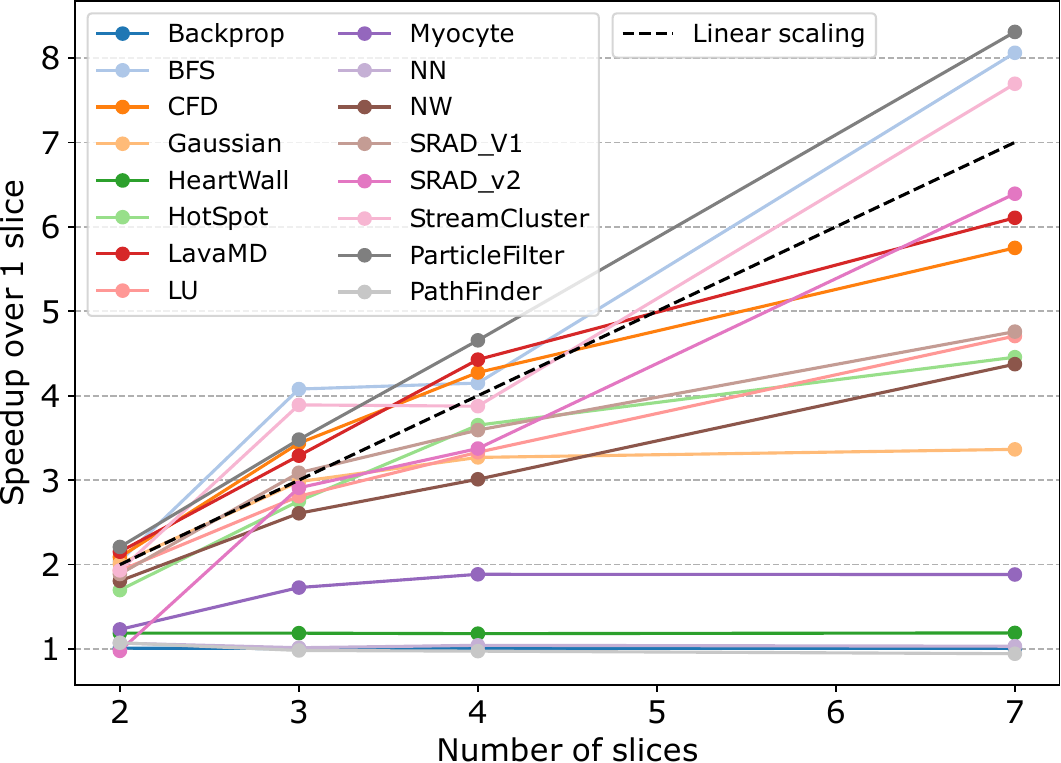}
\caption{MIG speedup of selected Rodinia benchmarks on a NVIDIA A100.}
\label{fig:MIG_scaling}
\end{figure}

In our experience with MIG, it is relatively common for applications to superscale. \Cref{fig:MIG_scaling} reports the scaling of a subset of the Rodinia kernels on an A100 according to the number of slices, many of which widely exceed linear scaling. Of course, there are also applications that scale approximately linearly or clearly sub-linearly from a certain instance size. Superscaling is probably more prevalent in MIG than in multicore processors as the memory and the bandwidth of the instances are isolated rather than shared, which favors great improvements for memory-bound kernels (e.g. {\em BFS} and {\em StreamCluster} superscale up to 7 slices but hardly improve from 3 to 4, as both sizes offer the same memory bandwidth).

From what we have discussed, to solve the problem, we will not assume the second point of the monotony hypothesis. However, we will assume the first point since we have rarely observed tasks taking longer when using more slices, and in the few cases that have occurred, the increase has been less than 2\%. 

\subsection{Instance isolation}\label{sec:instance_isolation}

As discussed, instance isolation is the distinguishing feature of MIG compared to other techniques for GPU task co-execution. Although NVIDIA claims that the isolation is total, an experimental verification is pertinent, as it is fundamental for the scheduling problem posed. In particular, it is necessary to verify that the execution of a task on an instance is independent of the tasks running simultaneously on other instances.

To achieve this, we run each of the 16 Rodinia kernels mentioned earlier on an instance of every possible size. First, the kernels are executed without any activity on the remaining slices, and then together with one or two randomly selected kernels. For example, we start by measuring the execution time of the {\em Backprop} kernel on a size-4 instance ${S_0, S_1, S_2, S_3}$  while keeping slices $S_4$, $S_5$ and $S_6$ idle. Next, we repeat the measurement but this time running a randomly chosen kernel simultaneously on the instance $\{S_4, S_5, S_6\}$. Then, {\em Backprop} is executed again on the size-4 instance, but now with two additional random kernels running on instances of sizes 2 and 1, assigned to $\{S_4, S_5\}$ and $\{S_6\}$, respectively. Finally, we repeat this for two random kernels, each using instances $\{S_4\}$ and $\{S_5\}$.

From this experiment, we calculate the deviation in {\em Backprop}'s execution times compared to its standalone execution on a size-4 instance. This process is repeated for {\em Backprop} across instances of sizes 3, 2, and 1. Subsequently, the entire procedure is performed for the other 15 kernels. We limit co-execution to 1 or 2 additional kernels and select them randomly to avoid excessive combinatorial complexity while ensuring the scenarios are diverse and representative.

\begin{table}[ht!]

\caption{Maximum percentage (\%) difference in execution time with MIG when running a kernel alone vs. co-running with other tasks on NVIDIA A100.}
\label{tab:isolation_test}
\centering
\small
\setlength{\tabcolsep}{6pt}
\begin{tabular}{lcccc}\toprule

 & \multicolumn{4}{c}{Instance size} \\ \cmidrule{2-5}
 Kernel & 1         & 2         & 3         & 4 \\ \midrule Backprop & 1.13         & 1.11         &  0.54        & 0.35\\
 Gaussian & 0.62        & 0.51         &  0.82        & 0.81\\
HotSpot& 1.48         & 1.96         & 1.72         & 0.83 \\
LavaMD & 0.34         & 0.41         & 0.23         & 0.23\\\bottomrule

\end{tabular}

\end{table}

\Cref{tab:isolation_test} shows the maximum percentage difference in execution time of some cores when running together with other applications compared to running them alone, for each instance size. For brevity, we have chosen 4 kernels instead of reporting all 16. Among them, we observe at most a 1.96\% maximum difference in {\em HotSpot} runtime when using 2 slices, which is in fact also the maximum value if we would report all 16 kernels.  In all cases we see very small time differences, which are probably more related to data caching between runs than to lack of isolation. These results demonstrate that the isolation provided by MIG is really effective.

\section{\AlgAbrev: Family of Allocations and Repartitioning\label{sec:algorithm}}

To solve the aforementioned problem, we propose \AlgAbrev, a three-phase algorithm for the MIG scheduling problem of a batch of tasks. The first phase addresses moldability by generating a family of allocations with the number of slices each task will use~(\Cref{alg_phase1}). In the second phase, instance repartitioning is used to solve the rigid problem associated with each allocation, determining the start time and specific slices to execute the tasks~(\Cref{alg_phase2}). Finally, the third phase refines the schedule by moving tasks from slices that reach the makespan to other slices (\Cref{alg_phase3}). An implementation of the algorithm is provided in \cite{far_implementation}, and \Cref{sec:alg_execution} briefly explains how it performs FAR output execution on a MIG-capable GPU.

\subsection{Phase 1: Generating a family of allocations}\label{alg_phase1}
To decide how many slices to assign to each task we adapt the method proposed by \cite{Turek92}, which generates a family of candidate allocations, assuming the first point of monotony explained in \Cref{sec:no-monotony}, but not the second.

The idea is to minimize the \textit{work} of the tasks, defined as the product of slices by the execution time, but not to make tasks too long by receiving few slices. As the work is the area that the task will occupy in the timetable, reducing it favors a smaller makespan. However, very long tasks can produce a large makespan by themselves. Therefore, the first allocation selects the minimum number of slices that minimizes the work of each task. Denoting with $a_i^1$ the number of slices assigned to the i-th task in the first allocation, we have that:
\begin{description}[itemsep=3pt]
    \item[\textbf{First allocation:}] $\quad a^1_i = \argmin\limits_{s \in C_G} \big(s \cdot t_i(s)\big),\quad 1 \leq i \leq n.$
\end{description}
The rest are built from the previous one by increasing the slices of the longest task while possible. That is, all tasks in allocation $a^{k+1}$ have the same assignment as in $a^k$, except for the longest one, which is increased to the next possible number of slices (if it cannot be increased, $a^k$ is the last allocation in the family):
\begin{description}
    \item [\textbf{Allocation (k+1)-th from the k-th:}]
    \begin{equation*}
        a^{k+1}_i =
        \begin{cases}
          \argmin\limits_{s \in C_G,\, s > a^k_i} \big(s \cdot t_i(s)\big) & \text{if } t_i(a^k_i) = \max\limits_{1 \leq j \leq n}  t_j(a^k_j), \\
          \quad a^k_i & \text{otherwise.}
        \end{cases}
    \end{equation*}
\end{description}

The amount of allocations in the family is $\mathcal{O}(|C_G| \cdot n)$. As $C_G$ is fixed for a GPU and has a small cardinal ($3$ for the A30 and $5$ for the A100/H100), the number of allocations is de facto $\mathcal{O}(n)$.

Finally, the following theorem, proved in \cite{Turek92}, bounds the error of some allocation of the family, with respect to the optimal solution of the moldable problem, using an algorithm A that solves the rigid problem. Thus, by applying A to all the allocations, and choosing the solution with the lowest makespan, we have a scheduler with guaranteed maximum error for the moldable problem. The requirement is that the makespan when solving the rigid problem of a fixed allocation is bounded by a combination of the total task area ($\sum_{i=1}^n s_i \cdot t_i(s_i)$) and the duration of the longest task ($h_{max}$). In successive phases, such an algorithm will be presented, and its makespan will be bounded in \Cref{sec:error_bounds} according to that combination, using this theorem to derive an approximation factor for the moldable version.



\begin{theorem}\label{th:preserve_upper_bound}
    Let A be an algorithm that solves the rigid (non-moldable) scheduling problem, producing a makespan $\omega$ bounded as follows:
    \begin{equation*}\label{eq:bound_form}
        \omega \leq \max\Big\{\alpha_1 \textstyle\sum\limits_{i=1}^n s_i\cdot t_i(s_i) + \beta_1\, h_{\text{max}}\,, ... ,\, \alpha_m \textstyle\sum\limits_{i=1}^n s_i\cdot t_i(s_i) + \beta_m\, h_{\text{max}} \Big\},
    \end{equation*}
    where $s_i$ is the number of slices given for the i-th task, $h_{\text{max}} = \max_i t_i(s_i)$, and $\alpha_i$, $\beta_i$ are coefficients. Then, there is at least one allocation $a^k$ of the family for which using A yields a makespan $\omega^k$ bounded as follows:
    \begin{equation*}\label{eq:approx_result}
        \omega^k \leq \max\{\alpha_1 \cdot \#slices_G + \beta_1, \dots, \alpha_m \cdot \#slices_G + \beta_m\} \cdot \omega^\star,
    \end{equation*}
    where $\omega^\star$ is the optimal makespan for the moldable problem.
    
\end{theorem}

\subsection{Phase 2: Instance scheduling by repartitioning}\label{alg_phase2}

The second phase of the algorithm is in charge of scheduling each of the allocations obtained in the previous phase, returning the one with the lowest makespan.

\begin{figure}[ht!]
\vspace{-3pt}
\centering
\includegraphics[width=0.8\columnwidth] {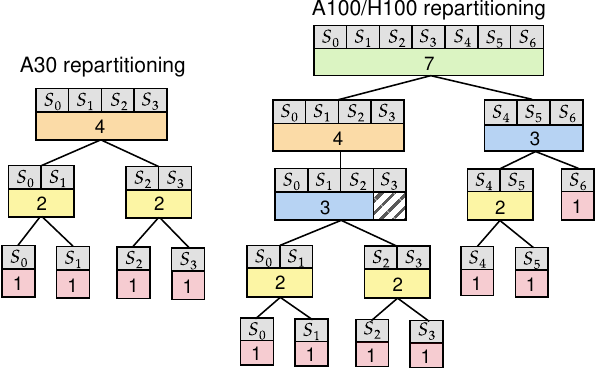}
\caption{Trees with instance repartitioning order for MIG-capable GPUs.}
\label{fig:MIG_repartition_trees}
\end{figure}

For each allocation, the algorithm follows a list scheduling strategy by trying to place tasks in instances that are repartitioned according to the trees of \Cref{fig:MIG_repartition_trees}. This method is outlined in \Cref{alg:repartitioning_method} (lines 7-16 for task placement and lines 17-23 for repartitioning). For example, for an A30 GPU, the algorithm first places consecutively the tasks assigned to its 4 slices; when there are no unscheduled tasks assigned to this size, it repartitions the instance into two instances of size 2 that are treated independently to place tasks assigned to this size; finally, each of these instances can be repartitioned into two instances of size 1 when there are no unscheduled tasks assigned to 2 slices. The process is similar for an A100/H100, but before repartitioning the instance with the first 4 slices, we try to use it for tasks assigned to only 3, due to the particularity explained in the introduction. In addition, the next instance to be used is the first one to be released (line 6), which self-balances the duration of the slices and makes it preferable to leave short tasks for the end; therefore, the tasks of each instance size are selected in LPT (Longest Processing Time first) order \cite{Graham1969} (lines 1-2).

The result is feasible, since the combinations of instances that could be generated (nodes from different branches at different levels of a tree) are precisely the valid MIG partitions shown in \Cref{fig:MIG_A30_A100_partitions}. The philosophy of MIG is that instances are independent and their dynamic repartitioning should not be conditioned by other instances. Thus, the existence of this tree and the feasibility of its combinations of disjoint instances should be fulfilled also in future GPUs, as we already explained in \Cref{sec:concepts_restriction}, so that the algorithm would be easily extrapolated and applicable.


\begin{algorithm}[ht!]

\SetKwComment{tcp}{\hfill \(\triangleright\)\,\,}{}
 \caption{Schedule an allocation by repartitioning}\label{alg:repartitioning_method}

\DontPrintSemicolon
  \KwData{Functions with task times $t_{i=1} ^n$ and creation/destruction times $t_{create}$/$t_{destroy}$, instance sizes $C_G$ and on allocation $a$}
  \KwResult{Instance tree of \Cref{fig:MIG_repartition_trees}, where each node has a list with the indexes, begin times, and durations of the tasks running on it}
  \BlankLine

  Group the task times by the number of slices $s \in  C_G$ given in $a$
  
  Order each group $s \in  C_G$ by decreasing time (LPT)
  
  $\text{{\em reconfig\_end}} := 0$
  
  Add the tree root instance $R$ of \Cref{fig:MIG_repartition_trees} to a min heap ordered by end time with: $R.end := 0, \, R.tasks := [\,]$
  
  \While{\textnormal{the heap is not empty}}{
    Pop from the heap the first instance $\mathcal{I}$ to end
    
    \tcp{Task placement}
    \If{\textnormal{there are unscheduled tasks assigned by $a$ to $|\mathcal{I}|$ slices}}{
    \tcp{Give time for $\mathcal{I}$ creation}
    \If{$\mathcal{I}$.tasks = [\,]}{
        $\text{{\em reconfig\_end}} := \max(\text{{\em reconfig\_end}},\, \mathcal{I}.end)$
        
        $\text{{\em reconfig\_end}} \pluseq t_{create}(|\mathcal{I}|)$
        
        $\mathcal{I}.end := \text{{\em reconfig\_end}}$
    }
    
    Take the longest unscheduled task $T_j$ using $|\mathcal{I}|$ slices
    
    \tcp{Execute $T_j$ right after in the same instance}
    
    $task := \big(index:= j,\,start := \mathcal{I}.end,\, time := t_j(|\mathcal{I}|)\big)$
    
    Append $task$ to $\mathcal{I}.tasks$
    
    $\mathcal{I}.end \pluseq t_j(|\mathcal{I}|)$

    Add $\mathcal{I}$ to the heap again
    }
    \tcp*[h]{Repartitioning}
    
    \ElseIf{\textnormal{there are unscheduled tasks}}{

    \tcp{Give time to destroy $\mathcal{I}$}
    
    \If{$\mathcal{I}$.tasks is not [\,]}{
        $\text{{\em reconfig\_end}} := \max(\text{{\em reconfig\_end}},\, \mathcal{I}.end)$
    
        $\text{{\em reconfig\_end}} \pluseq t_{destroy}(|\mathcal{I}|)$
    }

    \ForEach{\textnormal{instance child $C$ of $\mathcal{I}$}}{
    
        $C.end := \mathcal{I}.end$
        
        $C.tasks = [\,]$
        
        Add $C$ to the heap     
    }
    }
    
  }

\end{algorithm}


In each repartitioning decision we make sure that the first task of a new instance is not executed until the parent instance is destroyed and the new one is created, so that the periods of creation/destruction of instances are sequenced (this is what currently happens with MIG as we saw in \Cref{sec:concepts_restriction}). In \Cref{alg:repartitioning_method} this is done through the $\text{{\em reconfig\_end}}$ variable (line 3), which is global to the repartitioning. Although the creation/destruction periods are not explicitly returned, they are implicit in the output tree: the nodes of the tree that have allocated tasks will have to be created with the time indicated by the $t_{create}$ function (lines 8-11), and destroyed with the time indicated by $t_{destroy}$ if there is any descendant instance with allocated tasks (lines 18-20). A BFS traversal of the output tree allows to properly extract this reconfiguration info together with the obtained schedule.


In the worst case, LPT sorting (line 2) involves a complexity $\mathcal{O}(n\, log(n))$, which dominates over the task placement that requires $\mathcal{O}(n \cdot log(\#slices_G))$ to insert an instance in a heap with as many nodes as the binary tree of \Cref{fig:MIG_repartition_trees} for each task (line 16)\footnote{The height of the tree its $\leq \lceil log(\#slices_G) \rceil$ since $\#slices_G$ is halved at each level, so the number of nodes its $\leq \sum_{l=0}^{\lceil log(\#slices_G) \rceil} 2^l=2^{\lceil log(\#slices_G) \rceil + 1} - 1 = 2 \cdot 2^{\lceil log(\#slices_G) \rceil} -1 \leq 2 \cdot 2^{log(\#slices_G) + 1} = 4 \cdot \#slices_G$.}. Also, the cost of repartitioning decisions is $\mathcal{O}(\#slices_G)$ due to the number of nodes in a binary tree as in \Cref{fig:MIG_repartition_trees}\footnotemark[\value{footnote}]. Thus, the complexity of \Cref{alg:repartitioning_method} is $\mathcal{O}(n\, log(n) + \#slices_G)$, but since $\#slices_G$ is constant for a fixed GPU and small for current models, it is actually $\mathcal{O}(n\, log(n))$. Since the allocation family can have up to $\mathcal{O}(n)$ elements, executing the method on each one takes $\mathcal{O}(n^2\,log(n))$ time in the worst case. This cost is adequate for batches with tens or few hundreds of tasks that could accumulate in a real scenario. It must be noted that we control the batch formation criteria, and we could limit it by the number of accumulated tasks. In fact, in \Cref{sec:experimental} we will experimentally verify that with only a few tens of tasks the algorithm obtains very good results, and therefore the complexity is good enough.




The number of reconfigurations is limited by the number of non-leaf nodes in the tree, i.e., only 3 reconfigurations can occur in a schedule for A30, 7 reconfigurations for A100/H100, and at most $2 \cdot \#slices_G$ reconfigurations in the general case.\footnote{Since non-leaf nodes are at a tree level lower than $\lceil log(\#slices_G) \rceil$, their amount is $\leq \sum_{l=0}^{\lceil log(\#slices_G) \rceil -1} 2^l = 2^{\lceil log(\#slices_G) \rceil} -1 \leq  2 \cdot \#slices_G$.} Based on the reconfiguration times shown in \Cref{tab:reconfig_times}, its overhead should not be significant in a scenario with long tasks, especially if there are many of them. We will analyze this experimentally in \Cref{sec:experimental}.

In addition, the method can be used seamlessly for multiple A30s and multiple A100/H100s; for that, there would be as many trees as GPUs, and initially, there is one node for the root of each tree to start repartitiong on. In such a case, the time complexity would also  remain the same and the maximum number of reconfigurations would increase by a factor equal to the number of GPUs.

\subsection{Phase 3: Solution refinement}\label{alg_phase3}

\begin{figure}[ht!]
\centering
\includegraphics[width=\columnwidth] {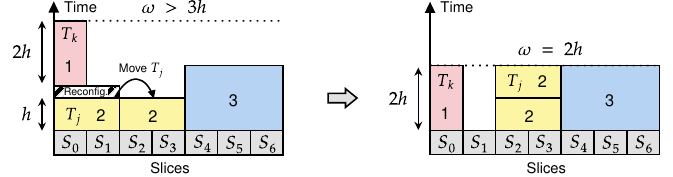}
\caption{Example of task movement with a makespan improvement of more than $3/2=1.5$ by moving the $T_j$ task to other slices.}
\label{fig:refinement_example}
\end{figure}

\textbf{Task movement.} 
To place a task, the previous phase considered the duration and location of those previously scheduled by using the first instance to be released, but not the duration of those that will be placed later on its slices. An example of its consequences is shown in \Cref{fig:refinement_example}, where task $T_k$ uses a slice previously occupied by $T_j$, and it is long enough so that $T_j$ should have been placed on other slices, even if they were not the first to be released. In our experience, this situation is relatively common and has a major impact in some cases (especially if there are few tasks in the batch), but can be easily mitigated by post-processing the solution to move some tasks. \Cref{fig:refinement_example} shows an example where moving $T_j$ reduces the makespan by a factor greater than $1.5$.

\begin{algorithm}[ht!]
\SetKwComment{tcp}{\hfill \(\triangleright\)\,\,}{}
 \caption{Schedule refinement}\label{alg:refinement}

\DontPrintSemicolon
  \KwData{Tree $S$ of \Cref{fig:MIG_repartition_trees} with the schedule resulting from phase 2}
  \KwResult{Tree $S$ with the schedule refined}
  \BlankLine

  Create an empty queue $Q$ for the open instance nodes
  
  $stop := false$ \quad\tcp*[h]{Can be used to limit iterations}

  $\omega := $ makespan of $S$
  
  \While{\textnormal{not $stop$}}{
    Push the leaf nodes of $S$ with slices reaching $\omega$ into $Q$
  
  \While{\textnormal{$Q$ is not empty}}{
    Pop the first instance node $\mathcal{I}$ of $Q$
    
    \If{$\mathcal{I}$ is the root node}{
        $stop := true$
        
        $break$
    }

    \tcp*[h]{Task movement}
    
     Get $\mathcal{I}^a \in S$ with $|\mathcal{I}^a| = |\mathcal{I}|,\,\mathcal{I}^a \neq \mathcal{I}$ for which the end time of its last slice $end(\mathcal{I}^a) =\max_{s \in \mathcal{I}^a} (s.end)$ is minimum

     Bin search task $T$ in $\mathcal{I}.tasks$ with $T.time < \omega - end(\mathcal{I}^a)$ which is closest to $(\omega - end(\mathcal{I}^a)) / 2$ 
    
     \If{\textnormal{task $T$ has been found}}{
                    Pop $T$ from $\mathcal{I}.tasks$
                    
                    Insert $T$ in $\mathcal{I}^a.tasks$ ordered by $T.time$
                    
                    Update end time of the slices $s.end$ for $s\in \mathcal{I} \cup \mathcal{I}^a$
            }

    \Else{
        \tcp*[h]{Task swapping}
        
        Find $T_k \in \mathcal{I}.tasks, T_j \in \mathcal{I}^a.tasks$ with: $0< (T_k.time - T_j.time) < (\omega - end(\mathcal{I}^a))$ and $(T_k.time - T_j.time)$ closest to $(\omega\!-\!end(\mathcal{I}^a))/2$

        \If{\textnormal{$T_k$ and $T_j$ have been found}}{
                    Pop $T_k$ from $\mathcal{I}.tasks$ and $T_j$ from $\mathcal{I}^a.tasks$
                    
                    Insert $T_k$ neatly in $\mathcal{I}^a.tasks$ and $T_j$ in $\mathcal{I}.tasks$
                    
                    Update end time of the slices $s.end$ for $s\in \mathcal{I} \cup \mathcal{I}^a$
        }
        \ElseIf{\textnormal{parent of $\mathcal{I}$ has not been added to $Q$}}{
            Push the node instance parent of $\mathcal{I}$ into $Q$ 
        }
    }
    
    }
    Update makespan $\omega$ from new slice endings

  }
    Traverse tasks lists of $S$, updating task start times and reconfig. times

\end{algorithm}


\Cref{alg:refinement} outlines the movement of tasks in the solution to improve scheduling in a low complexity manner. Iteratively, the algorithm searches for \emph{critical tasks}, i.e., using some slice that reaches the makespan, that can be moved to another instance without critical slices. To find such tasks the algorithm follows a greedy strategy by traversing the critical task subtree in a reverse BFS path (from the leaves towards the root). To do so, it starts with the leaves corresponding to the slices that reach the makespan (line 5), and expands them to their parents in case it is not possible to move any of their tasks (lines 23-24). For an instance $\mathcal{I}$ of a \emph{critical path} (i.e., in a sequence of critical tasks), it considers the alternative instance $\mathcal{I}^a$ with the same size (i.e., $\mathcal{I}^a \neq \mathcal{I}$ s.t. $|\mathcal{I}^a| = |\mathcal{I}|$) and minimum completion time for its slices (line 11). Then, among the tasks that are assigned to instance $\mathcal{I}$, it looks for the most suitable one that does not reach the makespan when moving it to $\mathcal{I}^a$ (line 12), that is, $T.time < \omega - end(\mathcal{I}^a)$ and $T.time$ closer to $\big(\omega - end(\mathcal{I}^a))/2$, where $\omega$ denotes the makespan. The reason for considering half of the available time is that it will be divided between $\mathcal{I}$ and $\mathcal{I}^a$ during the movement, and ideally, this division should be as balanced as possible. As the tasks of each node are sorted by duration, this can be done efficiently with a binary search. If successful, the algorithm inserts $T$ neatly into $\mathcal{I}^a$ (lines 13-16), and if not, it opens the parent instance of $\mathcal{I}$ unless it has been previously opened by another node (lines 23-24). The algorithm will have reduced the makespan of all the critical slices when all the nodes of an iteration are closed. Then, a new iteration will be started with the new critical slices (line 25 and again line 5). For each iteration the algorithm only recomputes the new makespan, so that the start times of the task that provide the detailed scheduling, and the reconfiguration times added/removed, are not computed until all the movements are done (line 26).

\begin{figure}[ht!]
\centering
\includegraphics[width=\columnwidth] {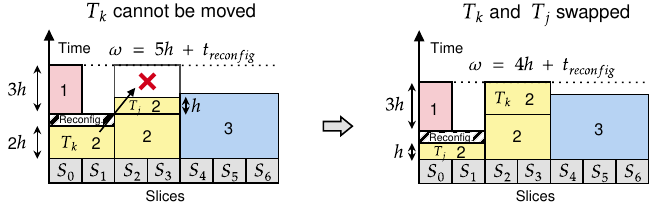}
\caption{Example of task swapping with a makespan improvement of almost $5/4=1.25$.}
\label{fig:refinement_swapping}
\end{figure}

\textbf{Task swapping.}
It may also happen that a critical task cannot be moved to another instance without reaching the current makespan, but can be swapped for a shorter task to improve it. This is illustrated in \Cref{fig:refinement_swapping} with an improvement factor of almost 1.25. Thus, if during an iteration of \Cref{alg:refinement} we cannot perform any movement from an instance $\mathcal{I}$ to its alternative $\mathcal{I}^a$, we look for a task $T_k$ of $\mathcal{I}$ and a task $T_j$ of $\mathcal{I}^a$ to swap them (lines 18-22). The requirements to improve with swapping are that $0 < t_k(s) - t_j(s) < \omega - end(\mathcal{I}^a)$, where $s = |\mathcal{I}| = |\mathcal{I}^a|$. However, we again try to choose the tasks so that $t_k(s) - t_j(s)$ is as close as possible to $(\omega - end(\mathcal{I}^a))/2$ to balance the time reduction in $\mathcal{I}$ with the increase in $\mathcal{I}^a$ in the best way according to the available margin.

\,
\textbf{Refinement stop and complexity.}
Obviously, the refinement process ends if the root node is opened in an iteration because the makespan of some slice could not be reduced (lines 8-10). As we will see in \Cref{sec:experimental}, it almost always ends that way after few moves/swaps, but it is still advisable to set a sufficiently strict stopping condition to avoid too long execution times in the general case. We propose to set a maximum number of iterations, or a minimum percentage of makespan improvement to avoid wasting time with too subtle refinements. In \Cref{sec:experimental} we experimentally analyze the improvements produced by refinement in the schedule.

It is easy to check that each iteration can be implemented in $\mathcal{O}(n \cdot \#slices_G)$ time complexity. This is because, at most, $\#slices_G$ critical tasks are moved or swapped (as much as initially opened critical slices) and the swapping requires $\mathcal{O}(n)$ time because the lists $\mathcal{I}.tasks$ and $\mathcal{I}^a.tasks$ are sorted, so we can efficiently search the tasks $T_j$ and $T_k$ that comply with line 18 of \Cref{alg:refinement} by a double-pointer strategy. In addition, the movement operation only requires $\mathcal{O}(log(n))$ time due to the task ordered insertion in its new instance. However, the actual cost will usually be much lower because not all tasks will be critical and not all will be allocated in a pair of instances.



\subsection{Scheduling execution}\label{sec:alg_execution}
Finally, to execute tasks on a MIG-capable GPU according to the schedule provided by the FAR algorithm, the nodes of the repartitioning tree (GPU instances) must be managed by different threads. As shown in \Cref{alg:GPU_exec}, the implementation described in \cite{far_implementation} traverses the repartitioning tree generated by \AlgAbrev in a descending manner. Each node with tasks calls to create its corresponding MIG instance on the GPU, sequentially executes the assigned tasks in that instance, and then destroys the instance. Additionally, every node --whether it has tasks or not-- requests its descendants to perform the same actions in separate threads. This approach enables tasks from different instances to run concurrently. \Cref{alg:GPU_exec} also reflects the calculation and reporting of the start and end time of each task, relative to the initial start of the batch execution. These times should be very similar to the ones used by FAR, due to the real isolation of the instances that we checked in \Cref{sec:instance_isolation} and the stability in the duration of reconfigurations and tasks that we also observed; in any case, this report allows verifying that the estimate of times simulated by FAR fits reality due to these hypotheses.

\begin{algorithm}[ht!]
\SetKwComment{tcp}{\hfill \(\triangleright\)\,\,}{}
\SetKwProg{Fn}{Function}{:}{}

 \caption{Scheduling execution on GPU}\label{alg:GPU_exec}

\DontPrintSemicolon
  \KwData{Repartitioning tree $S$ with the schedule produced by \AlgAbrev}
  \BlankLine

  Measure the current time as $init\_time$
  
  Call $execute\_tree$(root of $S$, $init\_time$) \quad\tcp*[h]{Recursive execution}

  \BlankLine

  \Fn{execute\_tree(\textnormal{tree instance} $\mathcal{I}$, $init\_time$)}{
  \tcp{Create, execute and destroy physically on GPU}
      \If{$\mathcal{I}$.tasks \textnormal{is not empty}}{ 
            $instance\_handler := GPU\_instance\_create(\mathcal{I})$
            
            \ForEach{\textnormal{task} $T$ \text{in} $\mathcal{I}$.tasks}{
                Show current time difference with $init\_time$ as $T$ start
                
                $GPU\_execute$($T$, $instance\_handler$)\;

                Show current time difference with $init\_time$ as $T$ end
            }
            $GPU\_instance\_destroy(instance\_handler)$    
      }

    \tcp{Repeat concurrently on children}
    
    \ForEach{\text{instance child} $C$ of $I$}{
        Run $execute\_tree(C, init\_time)$ in a new thread
    }

    Wait for all children threads to complete
  
  }
\end{algorithm}

  \vspace{-15pt}

\section{Multi-batch}\label{sec:multi_batch}


As seen, \AlgAbrev  solves the moldable offline scheduling for a single batch of tasks, but in a real scenario we deal with several batches over time; 
this implies the necessity of appending consecutive solutions in an intelligent way to maximize the resource usage.

\subsection{Concatenation issue} 
\begin{figure}[ht!]

\centering
\includegraphics[width=0.9\columnwidth] {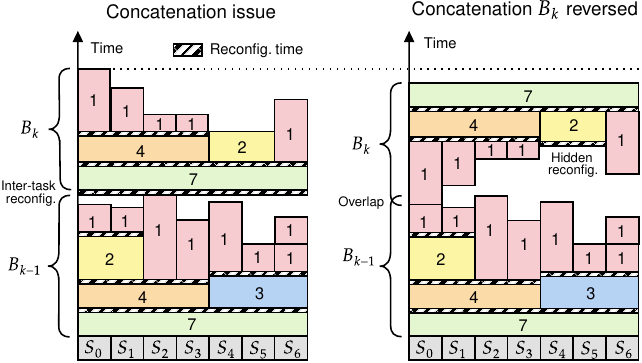}
\caption{Example of concatenation issue and improvement by reversing $B_k$.}
\label{fig:reversing_batch_concat}
\end{figure}

If we simply concatenate the scheduling produced by \AlgAbrev, the first task of a new batch typically waits until the last task of the previous batch finishes because \AlgAbrev schedules initial tasks using many slices (commonly all of them), and it is very likely that one of them will be the last slice to be released by the previous batch. For example, as shown in \Cref{fig:reversing_batch_concat} (left), a task that uses 7 slices in A100 will have been scheduled at the beginning of the scheduling of batch $B_k$ and cannot start until the last one of batch $B_{k-1}$ is finished since it will use its slices.

\subsection{Reverse batch scheduling to overlap} To mitigate this, our proposal reverses the order of schedules for consecutive batches. Tasks using fewer slices from one batch can be followed by similar tasks from the next batch, enhancing batch overlap. The schedule depicted in \Cref{fig:reversing_batch_concat} (right) shows the improvement when reversing the scheduling order of $B_k$ relative to $B_{k-1}$. Note that this overlap between batches refers to running tasks of the current batch $B_k$ in periods where its slices were idle in $B_{k-1}$; it does not mean to run tasks of both batches on the same slices at the same time. \Cref{fig:reversing_batch_concat} also shows that batch reversing can alleviate reconfigurations by hiding them in the overlap or eliminating them when the last instance of $B_{k-1}$ coincide with the first instance of $B_k$, which is more likely.

The best feasible overlap between the inverted schedule of $B_k$ and that of $B_{k-1}$ can be obtained in time $\mathcal{O}(\#slices_G)$ (constant time) by determining which slice forces $B_k$ to start later: just try to start $B_k$ after $B_{k-1}$ slice by slice and choose the option with the highest makespan (this is the best one that ensures feasibility on all slices).

\subsection{Moving/swapping tasks to overlap}

To increase the overlap between batches, we can move/swap tasks of the current batch $B_k$ to better fit with the schedule of the previous batch $B_{k-1}$. We focus on task in $B_k$ that starts immediately after the tasks in $B_{k-1}$, as these are critical for increasing inter-batch overlap.

\begin{figure}[ht!]
\centering
\includegraphics[width=0.9\columnwidth] {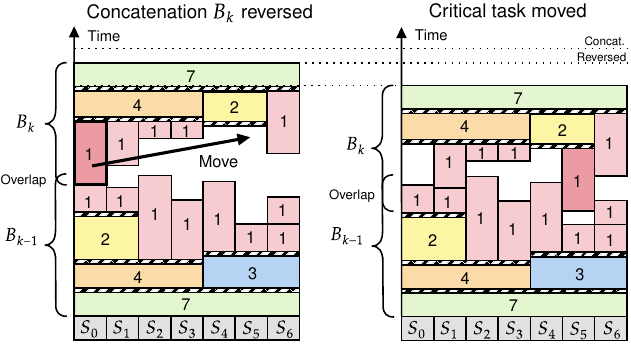}
\caption{Moving a critical task to improve the reversed batch concatenation.}
\label{fig:reloc_batch_concat}
\end{figure}

\begin{figure}[ht!]
\vspace{-3pt}
\centering
\includegraphics[width=0.9\columnwidth] {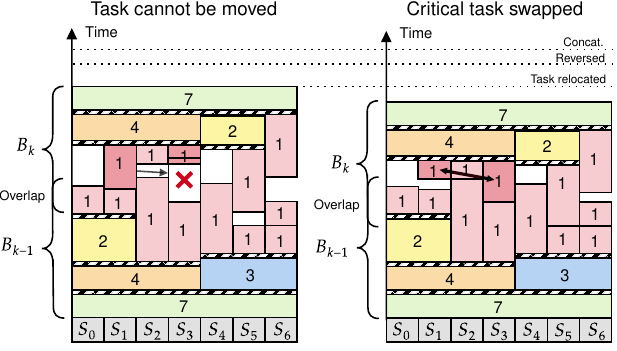}
\caption{Swapping a critical task which cannot be relocated to improve the reversed batch concatenation.}
\label{fig:swap_batch_concat}
\end{figure}

\begin{figure*}[ht!]
    \centering
    \resizebox{0.8\textwidth}{!}{ 
    \begin{minipage}{0.8\textwidth}
    \begin{subfigure}{0.3\textwidth}
        \includegraphics[width=\textwidth] {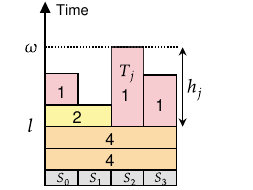}
        \caption{A30 possible solution.}
        \label{fig:MIG_bound_A30}
    \end{subfigure}
    \hfill
    \begin{subfigure}{0.33\textwidth}
        \centering
        \includegraphics[width=1\textwidth] {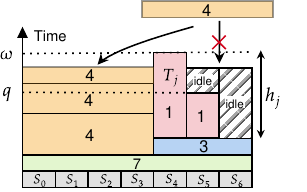}
        \caption{$S_5$ idle in A100/H100.}
        \label{fig:MIG_bound_s4-s5_idle}
    \end{subfigure}
    \hfill
    \begin{subfigure}{0.33\textwidth}
        \centering
         \includegraphics[width=1\textwidth]{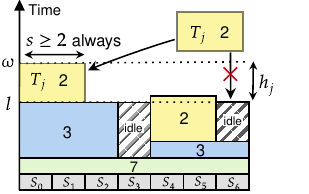}
        \caption{$S_6$ idle, but $S_4$ and $S_5$ not.}
        \label{fig:MIG_bound_s6_idle}
    \end{subfigure}
    \end{minipage}
    }
    \caption{Cases to deduce the upper bound of the approximation factor.}
\vspace{-5pt}
\end{figure*}

\Cref{fig:reloc_batch_concat} illustrates how the inverted concatenation from the previous example is further improved by moving a task from $B_k$, and \Cref{fig:swap_batch_concat} demonstrates how further improvement is possible by swapping tasks in $B_k$, even when no critical tasks can be moved. The operations are carried out in the same manner as the refinement described in \Cref{alg_phase3}, with the difference that the idle time of the slices between batches now takes the role that the makespan margin ($makespan - end(\mathcal{I}^a)$) played in the refinement. In addition, to perform the overlap increasing, the limiting slices must be recalculated after each iteration with complexity $\mathcal{O}(\#slices_G)$, but, as in the refinement, without having to update the start time of each task until the end of the iterations (see line 26 of \Cref{alg:refinement}).

\section{Approximation factor} \label{sec:error_bounds}

Next, we bound the makespan resulting from phase 2 of \AlgAbrev according to the form required by \Cref{th:preserve_upper_bound}, to obtain an upper bound of its factor with respect to the optimum of the moldable problem by applying the theorem. Note that phase 3 for refinement can only improve the solution, so that factor is also an upper bound if it is used. Reconfiguration times are not considered in this factor, but they must be negligible in relation to the duration of the tasks in the target scenario of the algorithm.

As proved in \cite{Shmoys1995}, concatenating schedules with an approximation factor $\rho$ has a {\em competitive ratio} $2\rho$ respect to the optimum, even if it is done without overlapping the executions. The {\em competitive ratio} is widely used in the literature to evaluate online schedulers \cite{Benoit2023}, and is defined as the maximum possible ratio between the makespan achieved by the scheduler processing the tasks on the fly and the optimal offline makespan if the tasks were available from the beginning in an offline way.

\subsection{Factor of NVIDIA A30}
For modeling the A30 it is sufficient to apply the List Scheduling proof to deduce the approximation factor \cite{Turek92}. For the sake of self-containment and since we will reuse the reasoning for the rest of the models, we briefly explain the particularized argument for this GPU.

The method can always use the released slices to immediately run other unscheduled tasks. This is because each node in its repartition tree has as ancestors all possible instance sizes larger than its own (\Cref{fig:MIG_repartition_trees}). Therefore, all tasks matched to larger sizes have already been scheduled and the remaining ones fit in its slices. As exemplified in \Cref{fig:MIG_bound_A30}, it means a schedule with the entire area occupied by tasks up to time $l$, when a critical task $T_j$ (one reaching the makespan $\omega$) starts.
It follows that the area up to $l$, which is $4\,l$ because A30 has 4 slices always occupied up to that point, is at most the total task area removing the $h_j = t_j(s_j)$ portion of $T_j$: $4\,l \leq \sum_{i=1}^n s_i \cdot t_i(s_i) - h_j$. Since also $h_j \leq h_{\max}=\max_{i=1} ^n t_i(s_i)$, it follows:
\begin{equation*}
    \omega = l+h_j \leq \tfrac{1}{4} \, \textstyle\sum\limits_{i=1}^n s_i \cdot t_i(s_i) - \tfrac{1}{4}\, h_j + h_j \leq \tfrac{1}{4} \textstyle\sum\limits_{i=1}^n s_i \cdot t_i(s_i) + \tfrac{3}{4} \,h_{\max}.
\end{equation*}
Now, we have the conditions to apply \Cref{th:preserve_upper_bound} deducing an approximation factor of $7/4\!=\!1.75$ respect to the optimal makespan of the moldable problem:
\begin{equation*}
    \omega \leq (\tfrac{1}{4} \cdot 4 + \tfrac{3}{4})\;\omega^\star = \tfrac{7}{4} \;\omega^\star.
\end{equation*}

If we use \AlgAbrev for two A30s instead of one, we have the same with 8 slices and can deduce that $\omega \leq 15/8\, \omega^\star = 1.875\,\omega^\star$ by the same argument. In general, if we use $g$ A30s, $\omega \leq (8\,g - 1)/(4\,g)\,\omega^\star$, which is bound by $2\,\omega^\star\; \forall g$.

Applying the result of \cite{Shmoys1995} for one A30 we get a competitive ratio of $2 \cdot 7/4 = 7/2 = 3.5$ for the tasks of all batches received online with respect to their offline optimum. In general, for any number of A30s we will have a factor of $2 \cdot 2 = 4$ for all batches.

\,

\subsection{Factor of NVIDIA A100/H100}\label{sec:approx_factor_A100_H100}

In this case, it is possible that the algorithm cannot place tasks on the slices of the first task to finish, not even partitioning it. For example, there may be unscheduled tasks requiring 4 slices and the first instance to finish is on slices $\{S_4, S_5, S_6\}$ (right branch of the tree), so no task can be placed and idle slices (a gap) are left. Following the same notation as in the previous section we will refer to it as idle period before time $l$, which is the start time of a critical task that reaches the makespan $\omega$. This can only happen in tree branches where not all instance sizes are present, since a node can be reached without having scheduled all tasks for larger instances. Therefore, it can only happen in the slices $\{S_4, S_5, S_6\}$, and in $S_3$ when we assign to the first 4 slices a task of size 3 (see \Cref{fig:MIG_repartition_trees}). For bounding we distinguish 3 cases:

\begin{enumerate}
    \item $\{S_4, S_5, S_6\}$ are not idle before time $l$.
    
    The only slice that can be idle before time $l$ is $S_3$. Therefore, the occupied area under $l$ is at least $6\,l$. Just as we reasoned for the A30 above, we have that $6\,l \leq \sum_{i=1}^n s_i \cdot t_i(s_i) - h_j$, with $h_j= \omega - l \leq h_{\max}$, so:
    \begin{equation*}
        \omega = l + h_j \leq \tfrac{1}{6} \textstyle\sum\limits_{i=1}^n s_i \cdot t_i(s_i) - \tfrac{1}{6} \, h_j + h_j \leq \tfrac{1}{6} \textstyle\sum\limits_{i=1}^n s_i \cdot t_i(s_i) + \tfrac{5}{6}\, h_{\max}.
    \end{equation*}
    \item $S_4$ or $S_5$ are idle before time $l$.
    
    When one of these slices is left idle, all unscheduled tasks need instances of size 4 because their tree branches have all the other sizes: 7, 3, 2, 1. The makespan is given by a task running in slices $S_0$ to $S_3$, or $S_4$ to $S_6$, so: $ \omega = \max(\omega_{S_0\!-\!S_3}, \omega_{S_4\!-\!S_6})$. In $S_0\!-\!S_3$ all tasks use 7 or 4 slices because, as we have explained, the possible unscheduled tasks are for 4 slices (the 4-slice node in the $S_0\!-\!S_3$ branch cannot have been partitioned). Therefore, the area under $\omega_{S_0\!-\!S_3}$ has no gaps and is bounded by all the task area: $4 \cdot \omega_{S_0\!-\!S_3} \leq \sum_{i=1}^n s_i \cdot t_i(s_i)$.
    
    On the other hand, all slices except perhaps $S_6$ are always busy below the first time $q$ when $S_4$ or $S_5$ are idle: as we argue, $S_0\!-\!S_3$ is filled with task using 7 or 4 slices with no gaps (\Cref{fig:MIG_bound_s4-s5_idle}). We use the same argument as before with $\omega_{S_4\!-\!S_6 - q \leq h_j \leq h_{\max}}$ for the $6$ slices always used below $q$, so $\omega_{S_4\!-\!S_6} \leq 1/6 \sum_{i=1}^n s_i \cdot t_i(s_i) + 5/6\,h_{\max}$. In short:
    \begin{gather*}
        \omega = \max(\omega_{S_0-S_3}, \omega_{S_4-S_6}) \\ 
        \leq \max \big(\tfrac{1}{4} \textstyle\sum\limits_{i=1}^n s_i \cdot t_i(s_i),\;\tfrac{1}{6} \textstyle \sum\limits_{i=1}^n s_i \cdot t_i(s_i) + \tfrac{5}{6}\,h_{\max} \big).
    \end{gather*}

    \item $S_6$ is idle before time $l$, but $S_4$ and $S_5$ not.
    
    Unscheduled tasks when $S_6$ is idle may need instances of sizes 2 or 4. If there are only tasks for size 4, we can use the argument of case 2 obtaining the same bound. If there are tasks for size $2$, $S_3$ may also be idle by passing through the node using only 3 slices of $S_0\!-\!S_3$. However, there are no tasks for size $1$, since there are none to place in $S_6$ and the parents of them in the tree have size 2, so there are tasks to place in 2. Therefore, the time $h_j$ of the critical task uses at least 2 slices and its area is at least $2h_j$ instead of $h_j$ (\Cref{fig:MIG_bound_s6_idle}). As there are always 5 slices occupied up to $l$ and exceeds an area of at least $2h_j$:
    \begin{equation*}
        \omega \leq \tfrac{1}{5} \, \textstyle\sum\limits_{i=1}^n s_i \cdot t_i(s_i) - \tfrac{2}{5} \, h_j + h_j \leq \tfrac{1}{5} \textstyle\sum\limits_{i=1}^n s_i \cdot t_i(s_i) + \tfrac{3}{5} \,h_{\max}.
    \end{equation*}
    
\end{enumerate}

Joining results, we have the following bound in the form of \Cref{eq:bound_form} and we can apply \Cref{th:preserve_upper_bound} to deduce a 2-approximation for the moldable problem:
\begin{gather*}
    \omega \leq \max \big(\tfrac{1}{6} \textstyle \sum\limits_{i} s_i\!\cdot \!t_i(s_i)\!+\!\tfrac{5}{6}\,h_{\max}, \frac{1}{4} \textstyle\sum\limits_{i} s_i\!\cdot\!t_i(s_i), \tfrac{1}{5} \textstyle\sum\limits_{i} s_i\!\cdot\! t_i(s_i)\!+\! \tfrac{3}{5} \,h_{\max}\big)\\
    \implies \omega \leq \max \big(\tfrac{1}{6} \cdot 7 + \tfrac{5}{6},\; \tfrac{1}{4} \cdot 7,\; \tfrac{1}{5} \cdot 7 + \tfrac{3}{5}\big)\, \omega^\star = 2 \, \omega^\star.
\end{gather*}

Applying the result of \cite{Shmoys1995} we get a competitive ratio of $2 \cdot 2 = 4$ for the tasks of all batches received online with respect to their offline optimum.

\section{Experimental evaluation\label{sec:experimental}}

For the experimental evaluation, we will employ the execution times obtained when 
using MIG on an A100 GPU for the Rodinia suite~\cite{Che2009}; however, due to the limited number of examples in the benchmark, and for the sake of generality, we also generated synthetic times mimicking the behaviors observed in those tests with some randomness. The logic of the task generator and the specific scenarios evaluated
are described in \Cref{sec:exp_times_generation}. All times, both measured and synthetically generated, are in seconds, to keep them in agreement with the reconfiguration times presented in \Cref{tab:reconfig_times}, which uses the \AlgAbrev algorithm.

\subsection{Experimental objectives}

Specifically, we develop a thorough evaluation of \AlgAbrev with 
a number of goals in mind, that are dissected throughout the section as
follows:

\begin{enumerate}[itemsep=2pt]
    \item Quantify the gap between simulated schedules generated by FAR (using its estimated times) and the real execution of these schedules on the GPU. Specifically, the goal of \Cref{sec:simulated_execution} is to verify if the simulation accurately reflects real-world performance, ensuring that it can be reliably used to evaluate the algorithm. This would allow it to be tested with synthetic workloads generated on demand, covering a wide range of scenarios.
    \item An estimation of the error of \AlgAbrev with respect to the 
    optimum for realistic workloads, comparing it with the theoretical 
    factor deduced in \Cref{sec:error_bounds}, is given in Section \Cref{sec:error_one_batch}.
    In addition, we analyse its error according to the number of scheduled tasks, 
    checking if it is necessary to schedule large task batches for good results.
    \item A comparison of \AlgAbrev versus alternative proposals such as fixed MIG 
    partitioning without reconfiguration and the only MIG scheduler found 
    in the literature \cite{Li2022} is given in \Cref{sec:exp_comparison_alternatives}.
    \item An analysis of the contribution of refinement techniques to improve scheduling and 
    its cost according to the number of iterations performed in this process is given
    in \Cref{sec:exp_refinement_contrib}.
    \item Finally, an evaluation of the extension of \AlgAbrev for multi-batch, 
    together with an analysis of the contribution of the improved batch combination is presented in \Cref{sec:exp_multibatch}.
\end{enumerate}

The results of \AlgAbrev shown include the reconfiguration time, which is limited and should not have a significant impact in a scenario with long tasks, as explained in \Cref{alg_phase2}. In particular, experiments will allow us to confirm this assumption, especially when comparing against a fixed partitioning without that overhead.

\subsection{Simulated execution}\label{sec:simulated_execution}

To evaluate the proposal in a wide range of scenarios, the task execution is simulated, but supported by real data. It should be noted that this simulation should reproduce such a simple process with high fidelity due to two validation keys: first, the verification of sequential reconfiguration behavior with highly stable real execution times used by \AlgAbrev, as detailed in \Cref{sec:reconfig_times}; and second, the confirmation of true instance isolation, as discussed in \Cref{sec:instance_isolation}. Under these assumptions, the transition from \AlgAbrev's output to the actual execution schedule is straightforward, simple and deterministic by the scheme explained in \Cref{sec:alg_execution}. In any case, it is convenient to corroborate the accuracy of this simulation by comparing with the real GPU scheduler provided \cite{far_implementation}.

\begin{table}[ht!]
\caption{Comparison of simulated vs. real end execution times, in a batch of 9 Rodinia kernels scheduled by FAR on NVIDIA A30.}
\label{tab:simulation_eval}
\centering
\small
\setlength{\tabcolsep}{8pt}
\begin{tabular}{lccc}\toprule
 & \multicolumn{2}{c}{End time (s)} &\\ \cmidrule{2-3}
 Kernel & Simulation         & Real execution         & Deviation (\%)\\ \midrule
PathFinder & 21.09 & 20.64 & -2.18
\\
LavaMD & 21.72 & 21.62 & -0.46
\\
HotSpot & 22.31 & 22.25 & -0.26
\\
Gaussian & 22.45 & 22.92 & 2.05
\\
NW & 22.79 & 23.26 & 2.02
\\
Huffman &  23.01 & 23.54 & 2.25
\\
HeartWall & 23.05 & 22.98 & -0.30
\\
ParticleFilter &  23.23 & 23.38 & 0.64
\\
\textbf{LU} & \textbf{28.86} & \textbf{28.24} & \textbf{-2.20}
\\\bottomrule
\end{tabular}
\end{table}

\Cref{tab:simulation_eval} presents the end times for the simulated and real executions of 9 Rodinia kernels scheduled in NVIDIA A30 as a batch using \AlgAbrev. The real execution times are derived from running the tasks on the GPU with the software we provide to the community \cite{far_implementation}. The end time refers to the total time from the start of the execution of all tasks until the completion of each task, which means that it represents partial durations of the entire execution, not the durations of the individual tasks. Along with the end times, the table shows the percentage deviation between the simulated and actual executions, providing a clear indication of their similarity. The critical task determining the makespan is highlighted in bold. For the sake of reproducibility, the scheduling software includes these 9 kernels as a pre-configured test, and the documentation clearly explains how to run it.

As can be seen in \Cref{tab:simulation_eval}, the deviations are minimal, with real execution times deviating at most 2\% from the simulated counterparts. This justifies that simulations based on the scheduler's timing parameters closely reflect real executions. This approach allows us to experiment with diverse on-demand workloads without the need to locate applications that exhibit these characteristics.

\subsection{Synthetic times generation\label{sec:exp_times_generation}}


%
As seen in \Cref{fig:MIG_scaling}, tasks initially scale linearly with the slice count and become sub-linear beyond a certain size. Also, some memory-bound tasks exhibit super-linear scaling initially. Our generator receives the number of tasks $n$, the percentage of tasks $p_s$ that scale well up to each instance size $s$ (near- or super-linear), and the percentage $p_{\text{sup}}$ that start with super-linear speedup from 1 to 2 slices since they are memory-bound. 
Also, we consider the transition of a memory-bound task to compute-bound with 0.3 of probability per slice.

For example, the times modeling for the A30 would be:
\begin{itemize}[itemsep=2pt]
     \item $n_1 = \lfloor n \cdot p_1 \rfloor$ tasks only exploit one slice\footnotemark{}, i.e., they have sub-linear speedup for all slice increments.
     \item $n_2 = \lfloor n \cdot p_2 \rfloor$ tasks exploit two slices\footnotemark[\value{footnote}]. Of them, $\lceil p_{\text{sup}} \cdot n_2 \rceil$ improve super-linearly from 1 to 2 slices (memory-bound), and $\lfloor (1-p_{\text{sup}}) \cdot n_2 \rfloor$ improve near-linearly (compute-bound). Both scale sub-linearly from 2 to 4 slices.
     \item $n_4 = \lfloor n \cdot p_4 \rfloor$ tasks exploit all slices\footnotemark[\value{footnote}]. On the one hand, $\lceil p_{\text{sup}} \cdot n_4 \rceil$ improve super-linearly from 1 to 2 slices, maintaining a super-linear speedup from 2 to 4 slices with a probability of $(1-0.3)^2$ (still memory-bound despite the increase of 2 slices), but moving to sub-linear speedup with a probability of $1-(1-0.3)^2$ (becomes compute-bound). On the other hand, $\lfloor (1- p_{\text{sup}}) \cdot n_4 \rfloor$ tasks improve near-linearly from 1 to 2 and from 2 to 4 slices.
\end{itemize}

\footnotetext{Applying percentages may result in non-integer values, so we floor the result. To create exactly $n$ tasks we iteratively add one to the size farthest from the exact: $n_j := n_j + 1$ for $j = \argmax_{i \in C_G}{(n \cdot p_i - n_i})$, while $\sum_{i \in C_G} n_i < n$.}

We generate the time $t_i(1)$ of each task on 1 slice with a uniform distribution $U(t_{\min}, t_{\max})$, where $t_{\min}$ and $t_{\max}$ are configurable parameters. Then, we generate the time $t_i(s+1)$ from $t_i(s)$ according to the speedup type as follows:%
\begin{gather*}
     t_i(s+1) = (s + r)\,/\,(s + 1) \cdot t_i(s), 
     \;\; \text{with } r \text{ taken random as } \\ r \in
     \begin{cases}
      N(-0.25, 0.25)_{[-0.5, 0]} & \text{for super-linear speedup},\\
      N(0.1, 0.1)_{[0, 0.2]} & \text{for near-linear speedup},\\
      N(0.75, 0.25)_{[0.5, 1]}& \text{for sub-linear speedup,}
    \end{cases}\\
    \begin{aligned}
    &\text{where } N(\mu, \sigma)_{[a, b]} \text{ is a normal distribution of mean } \mu \text { and}\\ &\text{standard deviation } \sigma \text{ clipped to } [a,b].
    \end{aligned}
\end{gather*}
%

\begin{figure}[t!]
    \centering
    \includegraphics[width=0.47\textwidth]{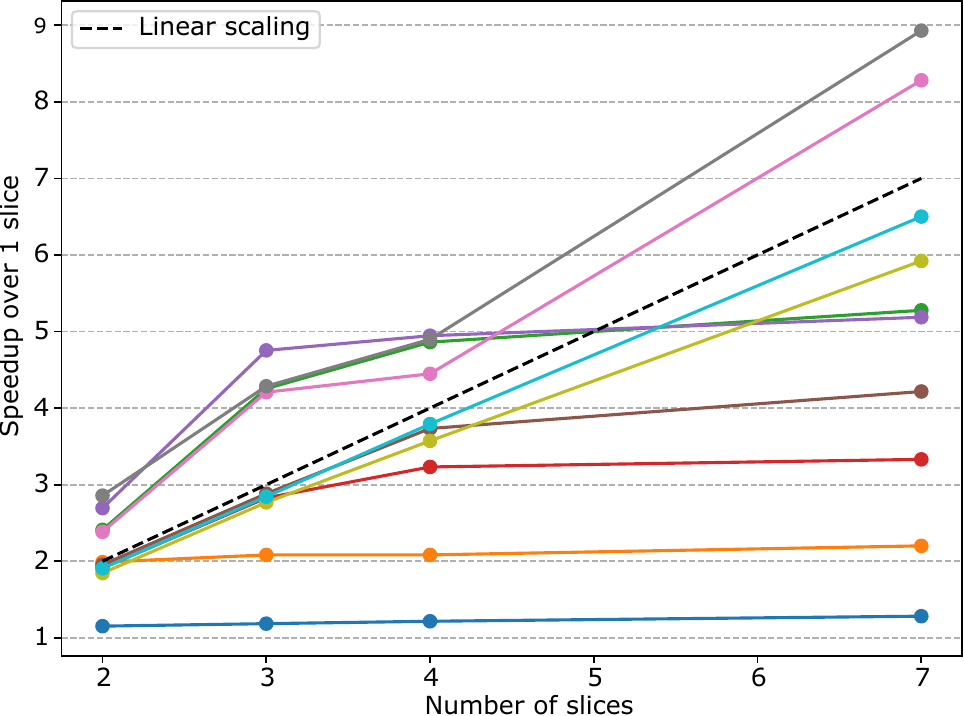}
    \caption{Synthetic speedups for $n\!=\!10$ tasks in A100/H100 using $p_1\!=\!p_2\!=\!10\%$, $p_3\!=\!20\%$, $p_4\!=\!p_7\!=\!30\%$ and $p_{\text{sup}}\!=\!50\%$.}
    \label{fig:synthetic_inputs_speedup}
\end{figure}

With $r\!=\!0$ the speedup would be linear and the higher $r$ the worse the improvement; therefore, we generate $r < 0$ for super-linear, $r \simeq 0$ for near-linear, and $r > 0$ for sub-linear speedup. 
We use normal distributions since application behaviors follow patterns and speedup factors that are unlikely to be uniformly distributed. Since the distribution is not bounded,
we clip the values to guarantee the desired speedup type (e.g., no $r > 0$ in a super-linear speedup case). In particular, we guarantee that $r \leq 1$, so that $t_i(s + 1) \leq t_i(s)$, fulfilling property 1 of monotony (\Cref{sec:no-monotony}).
Times for $s \notin C_G$ are useful for generation, but they are discarded afterwards as they are not input for the algorithm. \Cref{fig:synthetic_inputs_speedup} shows an example of generated speedups, which resemble those shown with Rodinia in \Cref{fig:MIG_scaling}.

Armed with the generator, we will employ synthetic loads with different combinations of parameters, covering several representative and diverse scenarios. In this sense, we denote the parameter configurations we will use for A100 as follows:

\begin{itemize}[itemsep=0.5pt]
    \item \POORSCALING: The percentages of tasks $p_s$ that scale well up to $s$ slices is set to $(p_1, p_2, p_3, p_4, p_7) = (50, 50, 0, 0, 0)$\%, i.e., half of the tasks can only exploit 1 slice well, the other half only scale well to 2 slices, and no task scales to more slices.
    \item \VARIEDSCALING: The scaling percentages are set to $\allowbreak (p_1, p_2, p_3, p_4, p_7) = (20, 20, 20, 20, 20)$\%, i.e., the same  20\% of tasks can get to exploit well to every possible instance size.

    \item \GOODSCALING: The task scaling percentages are set to $(p_1, p_2, p_3, p_4, p_7) = (0, 0, 0, 50, 50)$\%, i.e., half of the tasks that can exploit up to 4 slices well and the other half that can exploit all 7 GPU slices.

    \item \WIDETIMES: The range of task duration times is set to $[t_{\min}, t_{\max}] = [1, 100]$, i.e., a wide range between 1 and 100 seconds that conceives short tasks along with long duration tasks.

    \item \NARROWTIMES: The range of task duration times is set to $[t_{\min}, t_{\max}] = [90, 100]$, i.e., all tasks are long with a narrow range of times between 90 and 100 seconds.
\end{itemize}

For the different metrics that we will be presenting, the mean values of 1000 executions are reported; the deviations are not included for brevity, since they are relatively small in all cases and are not relevant for the discussion.

\subsection{Error with respect to the optimum}\label{sec:error_one_batch}

Since the optimal solution cannot be computed in a reasonable time (as it is an NP-hard problem), to evaluate \AlgAbrev we use the lower bound of the optimal makespan that arises from assigning minimum area to each task and distributing the total area evenly among the slices:

\[
baseline\!=\!\sum_{i=1}^n \min_{s \in C_G} (s \cdot t_i(s))\,/ \#slices_G \leq \omega^\star. 
\]

Results are presented hereafter as 

\[
\rho\!=\!\omega\, /\, baseline, 
\]
which overestimates the optimal makespan error: $ \omega \leq \rho\,\cdot\,\omega^\star$. 

\subsubsection{Comparison using Rodinia workloads}\label{sec:FAR_rodinia}

\AlgAbrev obtains $\rho\!=\!1.22$, i.e., at most 22\% error with respect to the optimal solution, using as input the times of the 16 Rodinia kernels profiled in the A100 (those shown in \Cref{fig:MIG_scaling}). This is a first indication that our proposal obtains good results even with reconfiguration overhead ({\em baseline} does not include it but \AlgAbrev makespan does); in fact, this value is much better than the theoretical worst case (approximation factor of 2 for the A100, i.e., 100\% of error). In this case, the execution time of \AlgAbrev has been between 1 and 2 milliseconds in several runs, confirming the efficiency of the algorithm. It should be noted that the tests have been performed on a personal machine that is not excessively powerful (AMD Ryzen 5 4600H processor).

\subsubsection{Comparison using synthetic workloads}\label{sec:FAR_synthetic_workloads}

For the sake of generality, \Cref{tab:ratios_error_A100} shows the $\rho$ value with synthetic inputs for A100, according to the number of tasks $n$, for the three task configurations according to their scalability described above. For the sake of brevity, only results for tasks of duration \WIDETIMES are reported, observing similar results for \NARROWTIMES times.

\begin{table}[ht!]
\setlength{\tabcolsep}{2pt}
\caption{Average of $\rho$ in 1000 execution of \AlgAbrev with \WIDETIMES for the A100.}
\label{tab:ratios_error_A100}
\centering

\fontsize{8pt}{8.4pt}\selectfont
\setlength{\tabcolsep}{5pt}
\renewcommand{\arraystretch}{1.2}
\begin{tabular}{lcccccc}\toprule
 & \multicolumn{6}{c}{Number of tasks ($n$)} \\ \cmidrule{2-7}
 Task configuration & 10         & 15         & 20         & 25         & 30         & 35\\ \midrule
\POORSCALING  & 1.23 & 1.08 & 1.04 & 1.03 & 1.02 & 1.02\\
\VARIEDSCALING                               & 1.20 & 1.08 & 1.04 & 1.03 & 1.02 & 1.02\\
\GOODSCALING                                  & 1.21 & 1.07 & 1.05 & 1.03 & 1.02 & 1.01\\ \bottomrule
\end{tabular}

\end{table}

\begin{figure*}[t!]
\centering
\includegraphics[width=0.9\textwidth]
{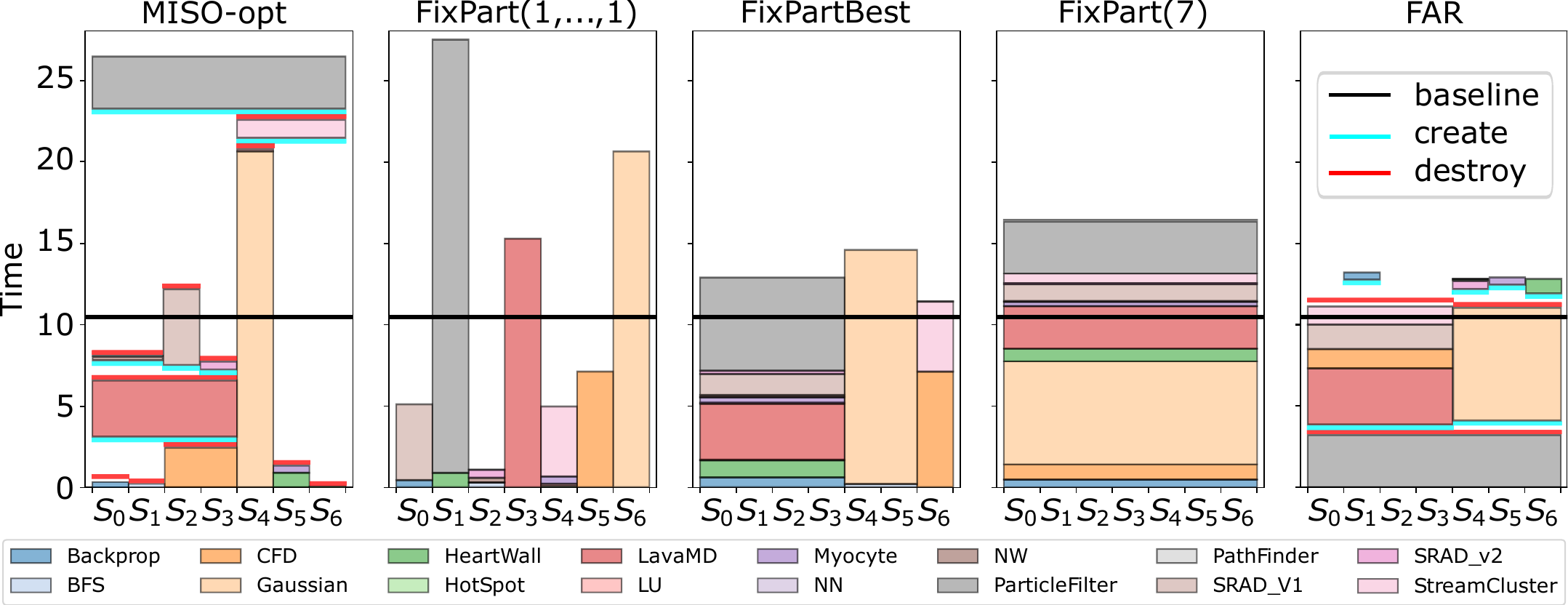}
\caption{Scheduling of the Rodinia kernels for NVIDIA A100 with different approaches.}
\label{fig:scheduling_comparison}
\end{figure*}

We observe negligible errors when using $n\! \geq \! 15$ tasks: below 10\% of error with 15 tasks, which progressively decreases to 2\% with 30-35 tasks. With just 10 tasks we get errors of around 20\% due to imbalances in the slice utilization of the last tasks to be placed, which has more impact with a reduced number of tasks. However, these errors are probably much smaller in reality since $\rho$ overestimates more with few tasks because the optimal schedule is further away from the baseline (the baseline is the perfect distribution of the task area among the slices). Also, the error for $n=15$ tasks, which is a similar amount to the Rodinia kernels used previously, is quite smaller because the execution times are longer and hence the reconfiguration overhead less significant. In all cases the errors are well below the approximation factor of 2 that was proved for the A100. Similar trends are observed for A30, but with much smaller errors for each $n$: $\rho \sim 1.01$ for $n\!=\!10$.

Additionally, for the same $n$ no significant differences are observed when varying task scaling capabilities: approximately the same error is made when scheduling a batch of tasks that can only leverage 1 and 2 slices (\POORSCALING), a batch of tasks that can leverage up to 4 and 7 slices (\GOODSCALING), and a very diverse batch of leverage capabilities evenly distributed among the slices (\VARIEDSCALING).

The average times to compute the schedule with the configurations in \Cref{tab:ratios_error_A100} are between 1 and 2 milliseconds, which is a negligible cost. The numbers of tasks reported in the table are those that can probably be accumulated to run \AlgAbrev. In any case, we have tested the time cost of the algorithm with $n=100$, $n=500$ and $n=1000$ tasks, obtaining average execution times of $12.32$ ms, $532.21$ ms and $1620.82$ ms respectively (1000 executions with \VARIEDSCALING configuration). This is still affordable, as with hundreds of tasks we are even below the time of a GPU reconfiguration (reported in \Cref{tab:reconfig_times}). Although a thousand tasks is probably excessive for a real scenario, it would still be equivalent to about 3 or 4 reconfigurations of the A100, which is adequate considering that we have posed an application scenario where the tasks last considerably longer than the reconfiguration time (otherwise the dynamism of MIG does not make much sense and it is preferable not to reconfigure).


\medskip

As a summary of conclusions: 

\begin{itemize}[itemsep=0.01pt]
    \item \AlgAbrev schedules close to the optimum ({\em baseline}) 
    and its error is actually well below the approximation factor. 
    \item The reconfiguration time is not relevant with long duration tasks. 
    \item It is not necessary to schedule a large number of tasks per batch to obtain 
    good results with long duration tasks. 
    \item The algorithm adapts well to different (synthetic) workloads in terms of the 
    GPU usage capabilities of the tasks.
    \item The cost of \AlgAbrev is affordable even for several hundred tasks.
\end{itemize}


\subsection{Evaluation of \AlgAbrev versus other approaches\label{sec:exp_comparison_alternatives}}

In the following, we compare \AlgAbrev against two alternative solutions:
\begin{itemize}[itemsep=3pt]
     \item \MISO: This is the solution proposed in~\cite{Li2022}. In this proposal, tasks are scheduled in FIFO order using the partition that maximizes the sum of speedups: the next partition will be $P\!=\!\{\mathcal{I}_0,\dots,\mathcal{I}_{\lvert P \rvert-1}\}\!\in\!\text{MIG-Part}_G$ maximizing $\sum_{i=0}^{\lvert P \rvert-1} \text{speedup}_{k+i}(\lvert\,\mathcal{I}_i\,\rvert )$, where $T_k$ is the next unscheduled task in FIFO order.
    
    \item \FIXPART($P$): Given a partition $P$, all tasks are scheduled in FIFO order on the first free instance. This represents a static configuration of the GPU done before the start of the execution. Note that, for the A100 and H100 GPUs, there are 19 different \FIXPART($P$) strategies (5 for the A30). We denote by \FIXPARTBEST the variant using the partition that provides the lowest makespan among all possible options.
\end{itemize}

\subsubsection{Comparison using Rodinia workloads}

\Cref{fig:scheduling_comparison} illustrates the scheduling comparison between those proposals and \AlgAbrev for the 16 Rodinia kernels. It highlights a major issue with \MISO, whose partitioning decisions are independent, resulting in longer idle periods as instances wait for preceding tasks to complete.
In fact, as a good sum of speeds is intended without considering the individual duration of the tasks, it is common that very long tasks run in parallel with very short ones and high idle periods arise between them (for example, {\em Gaussian} in \Cref{fig:scheduling_comparison}). This is solved in phase 1 of \AlgAbrev by using more slices to reduce the duration of the longest task in the successive candidate allocations. 

Also note that a fixed partition with 7 instances of size 1 generates a problem for long tasks that far exceed the ideal makespan for scheduling all tasks ({\em ParticleFilter}, {\em LavaMD} and {\em Gaussian} in \Cref{fig:scheduling_comparison}). Meanwhile, with 1 fixed instance of size 7, i.e., the equivalent of using the entire GPU for each task, there is also considerable room for improvement due to the poor scaling of some tasks to so many slices. Even with the best possible fixed partition (\FIXPARTBEST), there is room for improvement that \AlgAbrev exploits, which lies in the inflexibility to deal with tasks with different scaling capabilities without changing partitions. In fact, we find that these problems outweigh the sequential reconfiguration avoided by \FIXPART, even though the execution of these Rodinia kernels are not too long. Note that \AlgAbrev provides the least makespan schedule, even with idle periods due to sequential reconfiguration: the cyan/red rectangles of instance creation/destruction do not overlap leaving visible gaps.

Let $\sigma_{\text{A}}$ be the ratio between the makespan of an approach 
$A$ and the makespan of \AlgAbrev: %
$$\sigma_{\text{A}} = \omega_{A} / \omega_{\text{\AlgAbrev}}.$$

For the Rodinia kernels shown in \Cref{fig:scheduling_comparison}, we obtained $\sigma_{\MISO} = 2.10$, $\sigma_{\FIXPART(1,...,1)} = 2.18$, $\sigma_{\FIXPARTBEST} = 1.16$ and $\sigma_{\FIXPART(7)} = 1.26$. These results show an important advantage over these schedulers, consolidating the idea that task co-scheduling is worthwhile (\FIXPART(7) does not do it), betting on flexible dynamic reconfiguration despite the overhead due to partition changes (\FIXPART \ does not do it), and conditioning partitions to task execution times trying not to leave idle slices (\MISO does not do it).   

In terms of execution costs, there is not much to report compared to the other proposals. The scheduling calculation times for all of them are around the millisecond for the Rodinia subset, as we obtained for FAR in \Cref{sec:FAR_rodinia}.

\subsubsection{Comparison using synthetic workloads}

\Cref{tab:ratios_other_approaches} shows the $\sigma_A$ ratios for synthetic inputs based on 15 tasks over six workload types.

\begin{table}[ht!]
\centering
\setlength{\belowrulesep}{0pt}
\renewcommand{\arraystretch}{1.2}
\setlength{\tabcolsep}{2pt}
\caption{Average of $\sigma_A$ in 1000 executions with $n\!=\!15$ for A100.
}
\label{tab:ratios_other_approaches}


\fontsize{7.2pt}{8.76pt}\selectfont
\setlength{\tabcolsep}{2pt}
\begin{tabular}{lcccc}\toprule
& \multicolumn{4}{c}{Approach}         \\ \cmidrule{2-5}
Task configuration & \MISO & \FIXPART(1,...,1) & \FIXPARTBEST & \FIXPART(7) \\ \midrule
\POORSCALING, \NARROWTIMES    & 1.19  & 1.25        & 1.24    & 3.29  \\
\POORSCALING, \WIDETIMES     & 1.55  & 1.29        & 1.22    & 3.39  \\ \midrule
\VARIEDSCALING, \NARROWTIMES & 1.62  & 1.39        & 1.13    & 2.17  \\
\VARIEDSCALING, \WIDETIMES  & 2.03  & 1.47        & 1.09    & 2.16  \\ \midrule
\GOODSCALING, \NARROWTIMES    & 1.83  & 1.61       & 1.00    & 1.31  \\
\GOODSCALING, \WIDETIMES     & 2.14  & 1.78        & 1.01    & 1.28  \\ \bottomrule
\end{tabular}%

\end{table}

Note that \AlgAbrev improves more over \MISO when the range of times is wide (\WIDETIMES) versus more homogeneous task duration (\NARROWTIMES). This is because the coexistence of long and short tasks accentuates the idle gaps of that approach, but \AlgAbrev can handle with it. 

A great improvement over \FIXPART$(1,...,1)$ is observed when tasks are able to exploit up to 4 and 7 slices (\GOODSCALING), and analogously versus \FIXPART$(7)$ when they barely exploit 1 and 2 slices (\POORSCALING), which are obviously the worst cases for such proposals. However, \AlgAbrev also outperforms them by far for the opposite configurations, despite those proposals offer their best performance: around 30\% of improvement versus \FIXPART$(1,...,1)$ with tasks that barely scale (\POORSCALING), and also 30\% better versus \FIXPART$(7)$ with tasks that scale well (\GOODSCALING). In contrast, \FIXPARTBEST comes close to \AlgAbrev when the tasks exploit almost the entire GPU (\GOODSCALING), but is far below in the rest of the cases, which primarily motivates the use of MIG. Finally, \AlgAbrev improves substantially to all approaches when the GPU resource exploitation is mixed (\VARIEDSCALING), which is probably the most plausible case in real scenarios.

For the size of $n=15$ tasks reported in \Cref{tab:ratios_error_A100}, the cost of the different proposals have been very similar: in the case of \AlgAbrev each planning has taken on average 19.43 ms to compute, while the rest of the proposals move from 9.41 ms for \FIXPART$(1,...,1)$ to 13.45 ms for \MISO. These differences widen slightly if we increase the number of tasks but, as we already explained in \Cref{sec:FAR_synthetic_workloads}, the costs of \AlgAbrev are still quite affordable, especially if we consider the significant improvement obtained for the schedule. For example, for $n=1000$ tasks, \AlgAbrev took 1620.82ms to compute the schedule, \MISO took 1110.65 ms, \FIXPART$(1,...,1)$ took 935.65 ms, \FIXPARTBEST took 1035.26 ms and \FIXPART$(7)$ took 933.70 ms (average of 1000 runs with the \VARIEDSCALING configuration).


\medskip

In summary, \AlgAbrev improves significantly on both proposals by properly using GPU reconfiguration in very diverse scenarios. 
For example, as the last column of \Cref{tab:ratios_other_approaches} shows, if the tasks do not scale perfectly, \FIXPART(7) produces a makespan between 2 and 3 times larger than \AlgAbrev.
\AlgAbrev even improves significantly over the proposals without GPU reconfiguration in ideal cases such as \FIXPARTBEST, despite the reconfiguration costs that such proposals do not need. Only \FIXPARTBEST comes close to \AlgAbrev with tasks that can exploit almost all GPU resources, but it is an unremarkable case because the interest in using MIG arises from the inability of tasks to fully exploit the available resources. The cost of FAR is slightly higher than the other proposals, but the scheduling improvements make up for it. In fact, FAR is still very affordable even with many tasks.

\subsection{Refinement contribution\label{sec:exp_refinement_contrib}}

Another aspect to assess is the improvement provided by refinement (phase 3 of \AlgAbrev) in relation to its cost. It is intended to confirm that the movement/swapping of tasks raised for this phase has impact on the result without running too much operations. 

\Cref{tab:refinement_results} reports the percentage of improvement
\[
p_{\text{ref.}} = (\omega_{\text{no ref.}}/\omega_{\text{ref.}} - 1) \cdot 100,
\]
between the makespan without and with refinement for some of the workloads used before, along with the number of moves and swaps made until reaching an iteration where the method cannot improve. 

\begin{table}[ht!]

\caption{Average of $p_{\text{ref.}}$ (\%), along with the average number of movements and swaps operations performed in 1000 executions of refinement for A100.}
\label{tab:refinement_results}
\centering

%
%
%

\fontsize{6.7pt}{8.04pt}\selectfont
\setlength{\tabcolsep}{1.8pt}
\begin{tabular}{@{}lccccccccccc@{}}\toprule
 & \multicolumn{11}{c}{Number of tasks ($n$)} \\ \cmidrule{2-12}
 & \multicolumn{3}{c}{10} && \multicolumn{3}{c}{20} && \multicolumn{3}{c}{30}\\ \cmidrule{2-4}\cmidrule{6-8}\cmidrule{10-12}

 Task configuration & $p_{\text{ref.}}$ & moves & swaps && $p_{\text{ref.}}$ & moves & swaps && $p_{\text{ref.}}$ & moves & swaps \\
  \midrule
 \POORSCALING, \NARROWTIMES & 0.31 & 0.00 & 0.02 && 13.15 & 0.00 & 3.25 & &  11.45 & 0.00 & 4.26  \\
 \POORSCALING, \WIDETIMES  & 0.28 & 0.02 & 0.01 && 14.98 & 1.26 & 0.13  && 8.76 & 1.74 & 0.15\\ \midrule

 \VARIEDSCALING, \NARROWTIMES & 0.76 & 0.02 & 0.06 && 13.87 & 0.00 & 3.23  &&  9.04 & 0.00 & 4.31  \\ 
 \VARIEDSCALING, \WIDETIMES  & 3.21 & 0.17 & 0.06 && 11.45 & 1.14 & 0.09  &&  9.01 & 1.84 & 0.15 \\ \midrule

 \GOODSCALING, \NARROWTIMES & 0.78 & 0.01 & 0.01  && 13.44 & 0.00 & 3.01 &&  7.54 & 0.00 & 3.64  \\ 
 \GOODSCALING, \WIDETIMES  & 1.34 & 0.05 & 0.02 && 12.56 & 1.13 & 0.08  && 9.32 & 1.93 & 0.21 \\ \bottomrule
\end{tabular}

\end{table}

It can be observed in all cases a rather moderate number of moves and task swaps: at most around 1 move or 4 task swaps on average per refinement. As introduced in \Cref{alg_phase3}, the worst cost for a movement is $\mathcal{O}(log(n))$ and for a swap is $\mathcal{O}(n)$, so it can be intuited that the executions are not costly. In fact, the moderate costs for \AlgAbrev that we have reported in the previous sections, even for a large number of tasks, include the refinement phase. Nevertheless, significant improvements over scheduling without refinement are obtained, especially with $n=20$ tasks, where the percentage $p_{\text{ref.}}$ shows more than 10\% of improvement, but also with $n=30$ tasks, where $p_{\text{ref.}}$ drops slightly but also stays around that improvement value. For its part, with only $n=10$ tasks the refinement practically does not reduce the makespan, which may also explain the slightly higher errors we obtained for that number of tasks in \Cref{tab:ratios_error_A100} with respect to the ideal baseline. The conclusion is that refinement significantly and efficiently improves task scheduling for not very large batches, but if the number of tasks is too small it probably does not contribute anything. No intermediate or larger $n$ values have been included for brevity, but the trends are progressively maintained and stabilize at around $n=30$.

The different ranges for task duration in \Cref{tab:refinement_results} (\NARROWTIMES and \WIDETIMES) highlight the importance of considering both refinement operations: movement and swapping. It is observed that for a small range of long tasks (\NARROWTIMES) the operation that allows improvement is exclusively task swapping, but for a wide range including both long tasks and some short task (\WIDETIMES) the improvement comes primarily from movements. This is logical because with all long tasks it is very difficult to have idle time until the makespan of an instance to relocate a task of such duration (movement operation), but it will be likely to make a trade-off by gaining the difference in duration between the swapped tasks. When the duration of the tasks is heterogeneous, there are large idle gaps up to the makespan with respect to the duration of other tasks, which can be moved. On the other hand, the different scaling capabilities do not show great influence on the performance of the refinement, and simply corroborate their usefulness for diverse workloads.

\subsection{Multi-batch evaluation\label{sec:exp_multibatch}}
Finally, we evaluate the performance of our proposal for many task batches. We first focus on the contribution of the batch concatenation step presented in \Cref{sec:multi_batch}, and then on evaluating the algorithm as a whole for multi-batch task loads.

\subsubsection{Improved concatenation contribution}\label{sec:eval_improved_concat}

To evaluate the contribution of concatenation with overlap we take as a reference the trivial concatenation consisting of starting the scheduling of a batch just after the last task of the previous one. The improvement over it is disaggregated according to the contribution of batch reversal, and task move/swap operations performed after. 

\Cref{tab:concat_contribution} reports the improvement percentage 
\[
p_{\text{rev.}} = (\omega_{\text{trivial concat.}} / \omega_{\text{rev. concat.}} - 1) \cdot 100,
\]
between the makespan of the trivial concatenation and that obtained after task reversal, together with the percentage 
\[
p_{\text{move/swap}} = (\omega_{\text{trivial concat.}} / \omega_{\text{move/swap. concat.}} - 1) \cdot 100,
\]
between trivial concatenation makespan and the reversed one with also task moving/swapping. 

For each workload, a concatenation of 1001 different schedules produced by \AlgAbrev has been executed so that each one represents a batch. Therefore, we have 500 concatenations of a batch $B_k$ with a batch $B_{k+1}$ reverse and 500 concatenations where $B_{k+1}$ is not reverse as the one that is reversed is $B_k$. The mean improvement percentages are reported aggregated since no great difference was observed between the results of each type: practically the same when $B_{k}$ is the reversed one as when it is $B_{k+1}$. Moreover, in a real scenario, the number of tasks in each batch would be different, but for dimensionality issues and due to the analytical purposes of the experiment, all concatenated batches in each workload have the same number of tasks.

\begin{table}[h!]

\caption{Average of $p_{\text{rev.}}$ (\%) and $p_{\text{move/swap}}$ (\%) in the 1000 concatenations of 1001 \AlgAbrev schedules for A100.}
\label{tab:concat_contribution}
\centering

%
%
%
%
%
%
%
%
%

\fontsize{7.5pt}{9pt}\selectfont
\setlength{\tabcolsep}{2pt}

\begin{tabular}{lcccccccc}\toprule
 & \multicolumn{8}{c}{Number of tasks ($n$)} \\ \cmidrule{2-9}
 & \multicolumn{2}{c}{10}            && \multicolumn{2}{c}{20}&& \multicolumn{2}{c}{30}\\ \cmidrule{2-3}\cmidrule{5-6}\cmidrule{8-9}

 Task configuration & $p_{\text{rev.}}$ & $p_{\text{move/swap}}$ && $p_{\text{rev.}}$ & $p_{\text{move/swap}}$ && $p_{\text{rev.}}$ & $p_{\text{move/swap}}$ \\  \midrule
 \POORSCALING, \NARROWTIMES & 2.32 & 16.19 && 2.64 & 4.60 && 1.03  & 1.45  \\ 
 \POORSCALING, \WIDETIMES   & 3.23 & 14.47 && 3.82 & 4.10 && 0.98 & 1.01 \\ \midrule

\VARIEDSCALING, \NARROWTIMES & 5.43 & 16.22 && 4.53 & 4.82 && 0.86 & 1.01   \\
\VARIEDSCALING, \WIDETIMES  & 4.87 & 14.30 && 2.41 & 4.26 && 0.44 & 0.46   \\ \midrule

\GOODSCALING, \NARROWTIMES & 3.65 & 15.43 && 1.05 & 3.86 && 0.51 & 1.01    \\ 
\GOODSCALING, \WIDETIMES  & 5.54 & 13.12 && 2.03 & 3.76 && 0.23 & 0.30  \\ \bottomrule

\end{tabular}

\end{table}

\Cref{tab:concat_contribution} clearly shows that enhanced concatenation is especially useful versus trivial concatenation when batches have few tasks: $p_{\text{move/swap}}$ around 15\% improvement with $n=10$ tasks, which decays to just 4\% with $n=20$ tasks and disappears with $n=30$ tasks. This is logical given that: (1) the scheduling of each batch is closer to the ideal (perfectly balanced block) as its number of tasks increases and, therefore, a more basic concatenation has little room for improvement in such a case; (2) the relative weight in the total time of a move/swap of unbalanced tasks at the end of the schedule is lower when there are more tasks. In any case, the results show good performance for batches where \AlgAbrev has more error, i.e., those with very few tasks, and augur that the concatenation technique can mitigate it in the combination of schedules. In addition, we observe that the move/swap operations are much more important for improvement than the reversion of the batch to concatenate: $p_{\text{rev.}}$ 2-5\% of improvement with $n=10$, but $p_{\text{move/swap}}$ shows more than 10\% extra improvement than only reversing. However, as we saw in \Cref{sec:multi_batch}, the reversion has practically no computational cost, and moving/swapping has the same as we discussed in the refinement ($\mathcal{O}(log(n))$ the move and $\mathcal{O}(n)$ the swap). The rest of the task characteristics do not seem to have a clear influence on the results, and simply reaffirm the usefulness of the proposed concatenation for different scenarios.

We next analyze the cost and the particular utility of task movement or swapping, by the number of operations of each type that have been performed with each workload, as shown in \Cref{tab:moves_swaps_concat}. Again, it is the number of operations until an iteration where none can be performed to improve.

\begin{table}[h!]

\caption{Average number of movements or swaps in the 1000 concatenations of 1001 \AlgAbrev schedules for A100.}
\label{tab:moves_swaps_concat}
\centering

%
%
%
%
%
%
%
%
%
%
%

\fontsize{7.5pt}{9pt}\selectfont
\setlength{\tabcolsep}{2.5pt}
\begin{tabular}{lcccccccc}\toprule
 & \multicolumn{8}{c}{Number of tasks ($n$)} \\ \cmidrule{2-9}
 & \multicolumn{2}{c}{10} && \multicolumn{2}{c}{20} && \multicolumn{2}{c}{30}\\ \cmidrule{2-3}\cmidrule{5-6}\cmidrule{8-9}

 Task configuration & Moves & Swaps && Moves & Swaps && Moves & Swaps \\\midrule
 \POORSCALING, \NARROWTIMES & 0.48 & 9.72 && 0.04 & 1.32 && 0.01  & 0.82  \\
 \POORSCALING, \WIDETIMES  & 3.13 & 1.21 && 0.10 & 0.05 && 0.03 & 0.00 \\ \midrule

 \VARIEDSCALING, \NARROWTIMES & 0.30 & 9.52 && 0.02 & 1.12 && 0.00 & 0.62   \\
 \VARIEDSCALING, \WIDETIMES  & 3.51 & 1.12 && 0.13 & 0.06 && 0.05 & 0.01   \\ \midrule

 \GOODSCALING, \NARROWTIMES & 0.72 & 9.62 && 0.03 & 1.36 && 0.01 & 0.94    \\
 \GOODSCALING, \WIDETIMES  & 3.42 & 1.04 && 0.15 & 0.06 && 0.05 & 0.01 \\ \bottomrule
\end{tabular}

\end{table}

As for the refinement, \Cref{tab:moves_swaps_concat} shows that concatenating batches of homogeneous times (\NARROWTIMES) can only be improved in practice by task swaps, while concatenating batches of more varied duration (\WIDETIMES) improves mainly by task movement. We also notice more operations on average until no further improvement is possible compared to the refinement phase (see \Cref{tab:refinement_results}), but it still does not represent an excessive cost. The average time to compute each concatenation in this experiment was less than half a millisecond for $n=30$ tasks. Increasing the concatenation to batches of $n=1000$ tasks, an average time of 23.76 ms was obtained to compute each concatenation. In any case, we reiterate the option of limiting the number of operations when their improvement is low or too many have already been performed.

\subsubsection{Multi-batch performance} 

Finally, to gauge the overall error in multi-batch executions, we report a factor based on the competitive ratio, i.e., the maximum possible ratio between the makespan achieved by the online scheduler and the optimal offline makespan. Again, since it is not possible to compute the optimum in a reasonable time, we use the lower bound consisting in assuming that the minimum possible area of the tasks is evenly distributed among the slices\footnote{For brevity, in the single-batch scheduling context, the batch has been omitted from the notation of execution times. In the multi-batch context we use $t_i^B(s)$ to denote the execution time of the i-th task of batch $B$ using $s$ slices.}:
\[
baseline_{\text{multi-batch}}\!=\!\sum_{B \in Batches} \sum_{T_i \in B} \min_{s \in C_G} (s \cdot t_i^B(s))\,/ \#slices_G.
\]

That is, we report the improvement percentage:
\[
p_{\text{multi-batch}} = (\omega_{\text{multi-batch}}\,/\,baseline_{\text{multi-batch}} - 1) \cdot 100.
\]

\Cref{tab:ratios_error_multi_batch} reports $p_{multi-batch}$ for the same workloads used to obtain the $\rho$ ratios for a single batch in \Cref{tab:ratios_error_A100}. We use tens of thousands of tasks corresponding to 1001 batches to concatenate.  Again, we report only the results for \WIDETIMES, but the result of this experiment for \NARROWTIMES is similar. 

\begin{table}[ht!]
\setlength{\tabcolsep}{2pt}
\caption{Value of $p_{\text{multi-batch}}$ (\%) with \WIDETIMES and 1001 batches for A100.}
\label{tab:ratios_error_multi_batch}
\centering

\fontsize{8pt}{8.4pt}\selectfont
\setlength{\tabcolsep}{6pt}
\renewcommand{\arraystretch}{1.2}
\begin{tabular}{lcccccc}\toprule
 & \multicolumn{6}{c}{Number of tasks ($n$)} \\ \cmidrule{2-7}
 Task configuration & 10         & 15         & 20         & 25         & 30         & 35\\ \midrule
\POORSCALING  & 84.42 & 94.12 & 95.21 & 95.21 & 92.32 & 93.21\\
\VARIEDSCALING & 89.56 & 94.00 & 93.01 & 95.01 & 90.21 & 92.03\\
\GOODSCALING & 82.67 & 94.53 & 94.46 & 94.32  & 92.32 & 91.54\\ \bottomrule
\end{tabular}

\end{table}

By simply concatenating the batches we could expect up to twice the approximation ratio for single batch \cite{Shmoys1995}, which would mean a percentage of error between 102-116\% for $n \geq 15$ (see \Cref{tab:ratios_error_A100}). However, the multi-batch results in \Cref{tab:ratios_error_multi_batch} are 10-20\% below those values. For $n=10$, the results are up to 50\% below twice the single-batch error, as $baseline$ is not very tight for so few tasks. In contrast, $baseline_{\text{multi-batch}}$ does not exhibit this issue as it is calculated with the tasks from all batches. On the other hand, for larger numbers of tasks, the least tightly fitting may be $baseline_{\text{multi-batch}}$; the simplification made by both lower bounds comes from assuming that all tasks can be perfectly balanced with minimum area and no idle periods, but this incurs a larger error in multi-batch since it is assumed to be offline, but in reality tasks are received online. In any case, these results suggest good performance for multi-batch, although it is already justified by the tighter results reported for single-batch in \Cref{sec:FAR_synthetic_workloads}, and the improvement analysis with respect to trivial concatenation in \Cref{sec:eval_improved_concat}.

The scheduling cost is basically the sum of the times we have reported for FAR in \Cref{sec:FAR_synthetic_workloads}, and for the improved concatenation in \Cref{sec:eval_improved_concat}. As we have already discussed in those sections, they are good enough to support the efficiency of the proposal.

%
%

\section{Conclusions\label{sec:conclusions}}

In this paper we have motivated the opportunities of MIG by proposing a moldable scheduling problem to leverage this technology for task co-scheduling. Beyond presenting a complete scheduling proposal, accompanied by a useful C++ implementation on currently MIG-capable GPUs \cite{far_implementation}, this work highlights a promising line of research that can be greatly enriched by other scheduling approaches in the future. In addition, it suggests many metrics, workload characterizations, and evaluation techniques that may be useful for such future work.

In this work, we have considered a scenario with long running tasks in which batches of tasks are accumulated during the executions. For scheduling a batch we have presented \AlgAbrev, an efficient three-stage algorithm that combines several heuristics with an approximation factor of $1.75$ for NVIDIA A30 GPU and $2$ for the A100/H100 GPUs. In addition, we propose a method to concatenate the \AlgAbrev batch solutions, trying to overlap their executions to improve their combined makespan. 

Experimentally we have observed good solutions for real kernels profiled on the A100, with an error of less than 22\% of the optimum for a batch. In addition, we have obtained errors of less than 10\% on synthetic inputs that mimic real behaviors and cover a wide variety of workloads. These experiments have allowed us to verify that the overhead of MIG reconfiguration is not excessive if the scheduled tasks exhibit long durations, and in that case it is not necessary to schedule many tasks per batch. We observed a significant improvement over the only previous state-of-the-art approach (\MISO) and over a simple proposal which does not use dynamic MIG reconfiguration (\FIXPART). 

The third phase of \AlgAbrev significantly refines the solution for some batches, with improvements of around 15\% of the makespan for batches of 20 tasks. These improvements come from moving tasks when execution times are heterogeneous, and from swapping tasks when execution times are homogeneous. In addition, the proposed concatenation with overlap significantly 
improves multi-batch scheduling for batches with few tasks, where \AlgAbrev is a bit worse, achieving quite good overall performance and similar to larger batches.


For the future, we plan to leverage MIG in scenarios where tasks do not accumulate, for example because they are much shorter, and need to be scheduled {\em online}. We also consider extending the study for tasks with precedence constraints, deadlines, or different priorities. In some of these cases, GPU reconfiguration times may be more relevant and one of the challenge is to make the scheduler aware of them to avoid unhelpful reconfigurations.

\section*{Acknowledgements}
This work is funded by Grant PID2021-126576NB-I00 funded by MCIN/AEI/10.13039/501100011033 and by {\em ``ERDF A way of making Europe''}. 
We thank the HPC\&A group at Universitat Jaume I de Castell\'on for granting us access to A100 and H100 GPUs for profiling purposes.


\bibliographystyle{elsarticle-num}
\setlength{\bibsep}{0.2\baselineskip}
\bibliography{biblio}

\begin{thebibliography}{10}
\expandafter\ifx\csname url\endcsname\relax
  \def\url#1{\texttt{#1}}\fi
\expandafter\ifx\csname urlprefix\endcsname\relax\def\urlprefix{URL }\fi
\expandafter\ifx\csname href\endcsname\relax
  \def\href#1#2{#2} \def\path#1{#1}\fi

\bibitem{Abdelzaher1999}
T.~Abdelzaher, K.~Shin, Combined task and message scheduling in distributed
  real-time systems, IEEE Transactions on Parallel and Distributed Systems
  10~(11) (1999) 1179--1191.
\newblock \href {https://doi.org/10.1109/71.809575}
  {\path{doi:10.1109/71.809575}}.

\bibitem{Abreu2020}
L.~R. Abreu, J.~O. Cunha, B.~A. Prata, J.~M. Framinan, A genetic algorithm for
  scheduling open shops with sequence-dependent setup times, Computers \&
  Operations Research 113 (2020) 104793.
\newblock \href {https://doi.org/10.1016/j.cor.2019.104793}
  {\path{doi:10.1016/j.cor.2019.104793}}.

\bibitem{Allahverdi2022}
A.~Allahverdi, A survey of scheduling problems with uncertain interval/bounded
  processing/setup times, Journal of Project Management 7~(4) (2022) 255--264.
\newblock \href {https://doi.org/10.5267/j.jpm.2022.3.003}
  {\path{doi:10.5267/j.jpm.2022.3.003}}.

\bibitem{Allahverdi2015}
A.~Allahverdi, The third comprehensive survey on scheduling problems with setup
  times/costs, European Journal of Operational Research 246~(2) (2015)
  345--378.
\newblock \href {https://doi.org/10.1016/j.ejor.2015.04.004}
  {\path{doi:10.1016/j.ejor.2015.04.004}}.

\bibitem{Amarís2016A}
M.~Amarís, R.~Camargo, M.~Dyab, A.~Goldman, D.~Trystram, A comparison of gpu
  execution time prediction using machine learning and analytical modeling,
  2016 IEEE 15th International Symposium on Network Computing and Applications
  (NCA) (2016) 326--333\href {https://doi.org/10.1109/NCA.2016.7778637}
  {\path{doi:10.1109/NCA.2016.7778637}}.

\bibitem{Amoura2002}
A.~K. Amoura, E.~Bampis, C.~Kenyon, Y.~Manoussakis, Scheduling independent
  multiprocessor tasks, Algorithmica 32 (2002) 247--261.
\newblock \href {https://doi.org/10.1007/s00453-001-0076-9}
  {\path{doi:10.1007/s00453-001-0076-9}}.

\bibitem{Baker80}
B.~S. Baker, E.~G. Coffman, Jr., R.~L. Rivest, Orthogonal packings in two
  dimensions, SIAM Journal on Computing 9~(4) (1980) 846--855.
\newblock \href {https://doi.org/10.1137/0209064} {\path{doi:10.1137/0209064}}.

\bibitem{Bampis2002}
E.~Bampis, M.~Caramia, J.~Fiala, A.~V. Fishkin, A.~Iovanella, Scheduling of
  independent dedicated multiprocessor tasks, in: Algorithms and Computation,
  Springer Berlin Heidelberg, Berlin, Heidelberg, 2002, pp. 391--402.
\newblock \href {https://doi.org/10.1007/3-540-36136-7_35}
  {\path{doi:10.1007/3-540-36136-7_35}}.

\bibitem{Benoit2023}
A.~Benoit, L.~Perotin, Y.~Robert, H.~Sun, Online scheduling of moldable task
  graphs under common speedup models, in: Proceedings of the 51st International
  Conference on Parallel Processing, ICPP '22, Association for Computing
  Machinery, New York, NY, USA, 2023.
\newblock \href {https://doi.org/10.1145/3545008.3545049}
  {\path{doi:10.1145/3545008.3545049}}.

\bibitem{Blazewicz2006}
J.~Blazewicz, M.~Kovalyov, M.~Machowiak, D.~Trystram, J.~Weglarz, Preemptable
  malleable task scheduling problem, IEEE Transactions on Computers 55~(4)
  (2006) 486--490.
\newblock \href {https://doi.org/10.1109/TC.2006.58}
  {\path{doi:10.1109/TC.2006.58}}.

\bibitem{Blazewicz2001}
J.~Blazewicz, M.~Machowiak, G.~Mouni{\'e}, D.~Trystram, Approximation
  algorithms for scheduling independent malleable tasks, in: Euro-Par 2001
  Parallel Processing, Springer Berlin Heidelberg, Berlin, Heidelberg, 2001,
  pp. 191--197.
\newblock \href {https://doi.org/10.1007/3-540-44681-8_29}
  {\path{doi:10.1007/3-540-44681-8_29}}.

\bibitem{Bortfeldt2006}
A.~Bortfeldt, A genetic algorithm for the two-dimensional strip packing problem
  with rectangular pieces, E. Journal of Operational Research 172~(3) (2006)
  814--837.
\newblock \href {https://doi.org/10.1016/j.ejor.2004.11.016}
  {\path{doi:10.1016/j.ejor.2004.11.016}}.

\bibitem{Che2009}
S.~Che, M.~Boyer, J.~Meng, D.~Tarjan, J.~W. Sheaffer, S.-H. Lee, K.~Skadron,
  Rodinia: A benchmark suite for heterogeneous computing, in: 2009 IEEE
  International Symposium on Workload Characterization (IISWC), 2009, pp.
  44--54.
\newblock \href {https://doi.org/10.1109/IISWC.2009.5306797}
  {\path{doi:10.1109/IISWC.2009.5306797}}.

\bibitem{suscom_costero}
L.~Costero, F.~D. Igual, K.~Olcoz, Dynamic power budget redistribution under a
  power cap on multi-application environments, Sustainable Computing:
  Informatics and Systems 38 (2023) 100865.
\newblock \href {https://doi.org/10.1016/j.suscom.2023.100865}
  {\path{doi:10.1016/j.suscom.2023.100865}}.

\bibitem{Drozdowski1996}
M.~Drozdowski, Scheduling multiprocessor tasks — an overview, European
  Journal of Operational Research 94~(2) (1996) 215--230.
\newblock \href {https://doi.org/10.1016/0377-2217(96)00123-3}
  {\path{doi:10.1016/0377-2217(96)00123-3}}.

\bibitem{Du89}
J.~Du, J.~Y.-T. Leung, Complexity of scheduling parallel task systems, SIAM
  Journal on Discrete Mathematics 2~(4) (1989) 473--487.
\newblock \href {https://doi.org/10.1137/0402042} {\path{doi:10.1137/0402042}}.

\bibitem{Dutot2004}
P.-F. Dutot, L.~Eyraud, G.~Mouni{\'e}, D.~Trystram, Bi-criteria algorithm for
  scheduling jobs on cluster platforms, in: Proceedings of the sixteenth annual
  ACM symposium on Parallelism in algorithms and architectures, 2004, pp.
  125--132.
\newblock \href {https://doi.org/10.1145/1007912.1007932}
  {\path{doi:10.1145/1007912.1007932}}.

\bibitem{Fang2023}
J.~Fang, Y.~Rao, X.~Zhao, B.~Du, A hybrid reinforcement learning algorithm for
  2d irregular packing problems, Mathematics 11~(2) (2023).
\newblock \href {https://doi.org/10.3390/math11020327}
  {\path{doi:10.3390/math11020327}}.

\bibitem{Feitelson1997}
D.~G. Feitelson, L.~Rudolph, U.~Schwiegelshohn, K.~C. Sevcik, P.~Wong, Theory
  and practice in parallel job scheduling, in: Job Scheduling Strategies for
  Parallel Processing, Springer Berlin Heidelberg, Berlin, Heidelberg, 1997,
  pp. 1--34.
\newblock \href {https://doi.org/10.1007/3-540-63574-2_14}
  {\path{doi:10.1007/3-540-63574-2_14}}.

\bibitem{Floudas2005}
C.~A. Floudas, X.~Lin, Mixed integer linear programming in process scheduling:
  Modeling, algorithms, and applications, Annals of Operations Research 139
  (2005) 131--162.
\newblock \href {https://doi.org/10.1007/s10479-005-3446-x}
  {\path{doi:10.1007/s10479-005-3446-x}}.

\bibitem{Gomez-Villouta2010}
G.~G{\'o}mez-Villouta, J.-P. Hamiez, J.-K. Hao, Tabu search with consistent
  neighbourhood for strip packing, in: Trends in Applied Intelligent Systems,
  Springer Berlin Heidelberg, 2010, pp. 1--10.
\newblock \href {https://doi.org/10.1007/978-3-642-13022-9_1}
  {\path{doi:10.1007/978-3-642-13022-9_1}}.

\bibitem{Graham1969}
R.~L. Graham, Bounds on multiprocessing timing anomalies, SIAM Journal on
  Applied Mathematics 17~(2) (1969) 416--429.
\newblock \href {https://doi.org/10.1137/0117039} {\path{doi:10.1137/0117039}}.

\bibitem{Guerreiro2023}
R.~Guerreiro, A.~S. Santos, A.~Tereso, Online scheduling: A survey, in: 2023
  18th Iberian Conference on Information Systems and Technologies (CISTI),
  2023, pp. 1--6.
\newblock \href {https://doi.org/10.23919/CISTI58278.2023.10211826}
  {\path{doi:10.23919/CISTI58278.2023.10211826}}.

\bibitem{Harren2009}
R.~Harren, R.~van Stee, Improved absolute approximation ratios for
  two-dimensional packing problems, in: Approximation, Randomization, and
  Combinatorial Optimization. Algorithms and Techniques, 2009, pp. 177--189.
\newblock \href {https://doi.org/10.1007/978-3-642-03685-9_14}
  {\path{doi:10.1007/978-3-642-03685-9_14}}.

\bibitem{Hunold2015}
S.~Hunold, One step toward bridging the gap between theory and practice in
  moldable task scheduling with precedence constraints, Concurrency and
  Computation: Practice and Experience 27~(4) (2015) 1010--1026.
\newblock \href {https://doi.org/10.1002/cpe.3372}
  {\path{doi:10.1002/cpe.3372}}.

\bibitem{Jansen2002}
K.~Jansen, Scheduling malleable parallel tasks: An asymptotic fully
  polynomial-time approximation scheme, in: 10th Annual ESA, 2002, p.
  562–573.
\newblock \href {https://doi.org/10.1007/3-540-45749-6_50}
  {\path{doi:10.1007/3-540-45749-6_50}}.

\bibitem{Jansen2012}
K.~Jansen, A (3/2+$\epsilon$) approximation algorithm for scheduling moldable
  and non-moldable parallel tasks, in: Twenty-Fourth SPAA, 2012, p. 224–235.
\newblock \href {https://doi.org/10.1145/2312005.2312048}
  {\path{doi:10.1145/2312005.2312048}}.

\bibitem{Jansen2018}
K.~Jansen, F.~Land, Scheduling monotone moldable jobs in linear time, in: 2018
  IEEE international parallel and distributed processing symposium (IPDPS),
  IEEE, 2018, pp. 172--181.
\newblock \href {https://doi.org/10.1109/IPDPS.2018.00027}
  {\path{doi:10.1109/IPDPS.2018.00027}}.

\bibitem{Jansen2010}
K.~Jansen, R.~Th\"{o}le, Approximation algorithms for scheduling parallel jobs,
  SIAM Journal on Computing 39~(8) (2010) 3571--3615.
\newblock \href {https://doi.org/10.1137/080736491}
  {\path{doi:10.1137/080736491}}.

\bibitem{Johnston2018}
B.~Johnston, G.~Falzon, J.~Milthorpe, Opencl performance prediction using
  architecture-independent features, in: 2018 International Conference on High
  Performance Computing \& Simulation (HPCS), 2018, pp. 561--569.
\newblock \href {https://doi.org/10.1109/HPCS.2018.00095}
  {\path{doi:10.1109/HPCS.2018.00095}}.

\bibitem{Li2022}
B.~Li, T.~Patel, S.~Samsi, V.~Gadepally, D.~Tiwari, {MISO: exploiting
  multi-instance GPU capability on multi-tenant GPU clusters}, in: Proceedings
  of the 13th Symposium on Cloud Computing, 2022, p. 173–189.
\newblock \href {https://doi.org/10.1145/3542929.3563510}
  {\path{doi:10.1145/3542929.3563510}}.

\bibitem{Marchal2018}
L.~Marchal, B.~Simon, O.~Sinnen, F.~Vivien, Malleable task-graph scheduling
  with a practical speed-up model, IEEE Transactions on Parallel and
  Distributed Systems 29~(6) (2018) 1357--1370.
\newblock \href {https://doi.org/10.1109/TPDS.2018.2793886}
  {\path{doi:10.1109/TPDS.2018.2793886}}.

\bibitem{Mounie2007}
G.~Mounie, C.~Rapine, D.~Trystram, A $3/2$‐approximation algorithm for
  scheduling independent monotonic malleable tasks, SIAM Journal on Computing
  37~(2) (2007) 401--412.
\newblock \href {https://doi.org/10.1137/S0097539701385995}
  {\path{doi:10.1137/S0097539701385995}}.

\bibitem{mps}
NVIDIA, {Multi-Process Service}, URL \url{https://docs.nvidia.com/deploy/mps/}.

\bibitem{Perotin2024}
L.~Perotin, S.~Kandaswamy, H.~Sun, P.~Raghavan, Multi-resource scheduling of
  moldable workflows, Journal of Parallel and Distributed Computing 184 (2024)
  104792.
\newblock \href {https://doi.org/https://doi.org/10.1016/j.jpdc.2023.104792}
  {\path{doi:https://doi.org/10.1016/j.jpdc.2023.104792}}.

\bibitem{Perotin2021}
L.~Perotin, H.~Sun, P.~Raghavan, Multi-resource list scheduling of moldable
  parallel jobs under precedence constraints, in: Proceedings of the 50th
  International Conference on Parallel Processing, 2021, pp. 1--10.
\newblock \href {https://doi.org/10.1145/3472456.3472487}
  {\path{doi:10.1145/3472456.3472487}}.

\bibitem{Qi2024}
J.~Qi, W.~Xiao, M.~Li, C.~Yang, Y.~Li, W.~Lin, H.~Yang, Z.~Luan, D.~Qian,
  Elasticbatch: A learning-augmented elastic scheduling system for batch
  inference on mig, IEEE Transactions on Parallel and Distributed Systems
  35~(10) (2024) 1708--1720.
\newblock \href {https://doi.org/10.1109/TPDS.2024.3431189}
  {\path{doi:10.1109/TPDS.2024.3431189}}.

\bibitem{Shmoys1995}
D.~B. Shmoys, J.~Wein, D.~P. Williamson, Scheduling parallel machines on-line,
  SIAM Journal on Computing 24~(6) (1995) 1313--1331.
\newblock \href {https://doi.org/10.1137/S0097539793248317}
  {\path{doi:10.1137/S0097539793248317}}.

\bibitem{Steinberg97}
A.~Steinberg, A strip-packing algorithm with absolute performance bound 2, SIAM
  Journal on Computing 26~(2) (1997) 401--409.
\newblock \href {https://doi.org/10.1137/S0097539793255801}
  {\path{doi:10.1137/S0097539793255801}}.

\bibitem{Sudarsan2016}
R.~Sudarsan, C.~J. Ribbens, Combining performance and priority for scheduling
  resizable parallel applications, Journal of Parallel and Distributed
  Computing 87 (2016) 55--66.
\newblock \href {https://doi.org/10.1016/j.jpdc.2015.09.007}
  {\path{doi:10.1016/j.jpdc.2015.09.007}}.

\bibitem{Tan2021}
C.~Tan, Z.~Li, J.~Zhang, Y.~Cao, S.~Qi, Z.~Liu, Y.~Zhu, C.~Guo, {Serving DNN
  models with multi-instance GPUs: A case of the reconfigurable machine
  scheduling problem} (2021).
\newblock \href {https://doi.org/10.48550/arXiv.2109.11067}
  {\path{doi:10.48550/arXiv.2109.11067}}.

\bibitem{Tarplee2015}
K.~M. Tarplee, R.~Friese, A.~A. Maciejewski, H.~J. Siegel, Scalable linear
  programming based resource allocation for makespan minimization in
  heterogeneous computing systems, Journal of Parallel and Distributed
  Computing 84 (2015) 76--86.
\newblock \href {https://doi.org/10.1016/j.jpdc.2015.07.002}
  {\path{doi:10.1016/j.jpdc.2015.07.002}}.

\bibitem{Toksari2022}
M.~D. Toksar{\i}, G.~To{\u{g}}a, Single batch processing machine scheduling
  with sequence-dependent setup times and multi-material parts in additive
  manufacturing, CIRP Journal of Manufacturing Science and Technology 37 (2022)
  302--311.
\newblock \href {https://doi.org/10.1016/j.cirpj.2022.02.007}
  {\path{doi:10.1016/j.cirpj.2022.02.007}}.

\bibitem{Turek92}
J.~Turek, J.~L. Wolf, P.~S. Yu, Approximate algorithms scheduling
  parallelizable tasks, in: Proceedings of the Fourth Annual ACM Symposium on
  Parallel Algorithms and Architectures, 1992, p. 323–332.
\newblock \href {https://doi.org/10.1145/140901.141909}
  {\path{doi:10.1145/140901.141909}}.

\bibitem{far_implementation}
J.~Villarrubia, \href{https://github.com/artecs-group/FAR_MIG_scheduler}{Code
  repository: {FAR MIG} scheduler} (2025).
\newline\urlprefix\url{https://github.com/artecs-group/FAR_MIG_scheduler}

\bibitem{Wei2024}
X.~Wei, Z.~Li, C.~Tan, Optimizing gpu sharing for container-based dnn serving
  with multi-instance gpus, in: Proceedings of the 17th ACM International
  Systems and Storage Conference, SYSTOR '24, Association for Computing
  Machinery, New York, NY, USA, 2024, p. 68–82.
\newblock \href {https://doi.org/10.1145/3688351.3689156}
  {\path{doi:10.1145/3688351.3689156}}.

\bibitem{Wu2023}
X.~Wu, P.~Loiseau, Efficient approximation algorithms for scheduling moldable
  tasks, European Journal of Operational Research 310~(1) (2023) 71--83.
\newblock \href {https://doi.org/10.1016/j.ejor.2023.02.044}
  {\path{doi:10.1016/j.ejor.2023.02.044}}.

\bibitem{Ye2018}
D.~Ye, D.~Z. Chen, G.~Zhang, Online scheduling of moldable parallel tasks,
  Journal of Scheduling 21~(6) (2018) 647--654.
\newblock \href {https://doi.org/10.1007/s10951-018-0556-2}
  {\path{doi:10.1007/s10951-018-0556-2}}.

\bibitem{Zhang2024}
B.~Zhang, S.~Li, Z.~Li, Miger: Integrating multi-instance gpu and multi-process
  service for deep learning clusters, in: Proceedings of the 53rd International
  Conference on Parallel Processing, ICPP '24, Association for Computing
  Machinery, New York, NY, USA, 2024, p. 504–513.
\newblock \href {https://doi.org/10.1145/3673038.3673089}
  {\path{doi:10.1145/3673038.3673089}}.

\end{thebibliography}






\end{document}